# Orbital hybridization in graphene-based artificial atoms

Yue Mao[1,§], Hui-Ying Ren[2,3,§], Xiao-Feng Zhou[2,3,§], Hao Sheng[2,3], Yun-Hao Xiao[2,3], Yu-Chen Zhuang[1], Ya-Ning Ren[2,3,*], Lin He[2,3,*], and Qing-Feng Sun[1,4,*]

**Affiliations:**
[1]International Center for Quantum Materials, School of Physics, Peking University, Beijing, 100871, China
[2]Center for Advanced Quantum Studies, School of Physics and Astronomy, Beijing Normal University, Beijing, 100875, China
[3]Key Laboratory of Multiscale Spin Physics, Ministry of Education, Beijing, 100875, China
[4]Hefei National Laboratory, Hefei 230088, China
[§]These authors contributed equally to this work.
*Correspondence and requests for materials should be addressed to Ya-Ning Ren (e-mail: yning@mail.bnu.edu.cn), Lin He (e-mail: helin@bnu.edu.cn), and Qing-Feng Sun (e-mail: sunqf@pku.edu.cn).

**Intraatomic orbital hybridization and interatomic bond formation are the two fundamental processes when real atoms are condensed to form matter[1,2]. Artificial atoms mimic real atoms by demonstrating discrete energy levels attributable to quantum confinement[3–8]. As such, they offer a solid-state analogue for simulating intraatomic orbital hybridization and interatomic bond formation. Signatures of interatomic bond formation has been extensively observed in various artificial atoms[9–17]. However, direct evidence of the intraatomic orbital hybridization in the artificial atoms remains to be experimentally demonstrated. Here we, for the first time, realize the orbital hybridization in artificial atoms by altering the shape of the artificial atoms. The anisotropy of the confining potential gives rise to the hybridization between quasibound states with different orbital quantum numbers within the artificial atom. These hybridized orbits are directly visualized in real space in our experiment and are well reproduced by both numerical calculations and analytical derivations. Our study opens an avenue for designing artificial matter that cannot be accessed on real atoms through experiments. Moreover, the results obtained inspire the progressive control of quantum states in diverse systems.**

When atoms condense to form matter, two essential processes emerge: i) interatomic attractive forces bind atoms together, stabilizing them through energy loss, known as chemical bonds; ii) intraatomic potential fields become anisotropic, facilitating the recombination of atomic orbitals with close energy, referred to as orbital hybridization[1,2]. Confinement within quantum dots (QDs) leads to a series of bound states analogous to electron orbitals in real atoms[3–8], rendering QDs as artificial atoms[5,6]. Over decades, extensive research has focused on the coupling between these artificial atoms to simulate and explore the formation of matter at the atomic level. The artificial-atomic analogy of bond formation, as one essential process of real matter formation, has been widely investigated across various types of artificial atoms, including semiconductor-based structures[9–13], metal surfaces[15], and graphene QDs[14,16,17]. However, direct experimental evidence of the other essential process in the formation of matter—namely, intraatomic orbital hybridization—remains a significant challenge and has yet to be realized in artificial atom.

Here, we realize the orbital hybridization in graphene QD, a well-studied artificial atom, by introducing anisotropy of confining potentials. By altering the shape of QDs from circular to elliptical, the anisotropy of the confining potential induces the orbital hybridization between quasibound states of s-orbital and d-orbital. The hybridized orbitals are directly visualized in real space through scanning tunneling microscopy (STM) measurements, and these findings are well corroborated by both numerical calculations and analytical derivations. When progressively increasing the anisotropy of the QDs, our experimental results supported by theoretical calculations demonstrate that the energy splitting between the hybridized states increases, highlighting the enhancement of the orbital hybridization.

**Physical pictures**

In an isolated real atom, electrons experience an isotropic confining potential from the atomic nucleus. Due to this isotopy, electron orbitals are formed and characterized by the orbital quantum number $m$. Specifically, orbitals with $m=0$ and $m=1$ correspond to the s and p orbitals, respectively, as illustrated in Fig. 1a. However, when atoms condense to form matter, the isotropic nature of the electron potential is typically disrupted by neighboring atoms. For instance, in graphene, the equipotential line of electrons can be triangularly shaped by the influence of the three nearest atoms, as depicted in Fig. 1b. Consequently, the $s, p_x, p_y$ orbitals, which are close in energy, hybridize to form three superimposed states[1,2], a phenomenon well-known as $sp^2$ hybridization.

In analogy, QDs have a confining potential that also results in discrete energy levels for electrons[3,4]. In principle, it is feasible to achieve orbital hybridization when the rotation invariance of the confining potential is broken. We here focus on exploring the orbital hybridization in graphene QDs[14,16–36] due to the capability for in-situ creation and manipulation in our experimental setup[14,17,28,30–32]. Specifically, we investigate circular and elliptical QDs, schematically depicted in Figs. 1c and 1d, to elucidate the effects of anisotropic confining potential on orbital hybridization. The circular QD possesses an isotropic confining potential that preserves rotational invariance, allowing its quasibound states to be characterized by (m, n), with $m$ the orbital quantum number and $n$ the principal quantum number (see Supplementary information for wavefunction $\psi_{(m,n)}$ of the circular QD). Typically, the quasibound states with the lowest and the second lowest energies are respectively (m, n)=(0, 1) and (1, 1) labelled as *s1* and *p1*[22,25,26,30]. The energy disparity between these two states is significant, and they are also notably distant from other states in energy, thereby exhibiting almost no orbital hybridization (see Extended data Fig. 1). The subsequent two states are (m,n)=(0, 2) and (2, 1) and labelled by *s2* and *d1*, respectively[22,25,26,30]. The local density of states (LDOS) of the two states (0, 2) and (2, 1), obtained consistently from both numerical calculation and analytical wavefunction (details and comparison between different methods[20] are provided in Supplementary information and Figs. S1-S3), display rotation symmetry as depicted in Fig. 1e. Therefore, although the two states are usually very close in energy, there is no hybridization between these states within the circular QD.

In the elliptical QD, the rotation symmetry is broken, allowing the hybridization between the two states (0, 2) and (2, 1). Consequently, they recombine to form new states:

$$\psi_{sd+} = \psi_{(0,2)} + \alpha\psi_{(2,1)}, \qquad \psi_{sd-} = -\alpha\psi_{(0,2)} + \psi_{(2,1)}, \qquad (1)$$

where $\alpha \in [0,1]$ is the recombining coefficient indicating the hybridization strength. (Here the normalization factor $1/\sqrt{1+|\alpha|^2}$ has been omitted and the coefficient $\alpha$ can be complex number in principle, see Supplementary information and Fig. S4 for the details). Analogous to real atomic orbitals, the states (0, 2) and (2, 1) possess orbital numbers $m = 0$ and $m = 2$, and thus can be regarded as orbitals s and d, respectively. In this sense, this hybridization can be called "sd hybridization". In Fig. 1f, we show the consistent LDOS of hybridized states obtained from numerical calculation and analytical wavefunction. Because the orbital numbers of (0, 2) and (2, 1) differ by 2, the sd hybridization is induced by the elliptical potential (details and comparison between different methods[20] provided in Supplementary information and Fig. S5). The hybridized states $\psi_{sd+}$ and $\psi_{sd-}$ exhibit $\pi$-period versus polar angle, with a $\theta$ shape and a rotated $\theta$ shape respectively.

**Experimental observation**

To experimentally explore the orbital hybridization induced by anisotropic confining potentials, we systematically study the quasibound states generated by graphene QDs with various asymmetric potentials. These QDs are formed by inserting an interfacial monolayer 1T'-phase transition-metal dichalcogenide (TMD) nanoscale island into a graphene/2H-phase TMD heterostructure, as schematically depicted in the bottom panel of Fig. 2a. The 1T'-phase and 2H-phase TMD introduce different electronic doping in the topmost graphene layer, as shown in the top panel of Fig. 2a, creating high-quality p-n junctions with Coulomb confining potentials[30–32], i.e., the QDs. These QDs confine the massless Dirac fermions in graphene[17,30,31]. In our experimental setup, high-quality graphene/TMD heterostructures are achieved using transfer technology of graphene monolayer onto two different mechanically exfoliated TMD sheets $WSe_2$ and $MoSe_2$. The results obtained from these two different heterostructures are quite similar (see Methods section for details). STM tip pulses are employed to create an interfacial nanoscale monolayer TMD island and trigger its phase transition (from the H phase to the T' phase) at the heterostructure interface[17,30,31,37–40]. This method allows for the in-situ tailoring of the geometry of the interfacial TMD island using the STM tip[32,41,42], thereby enabling precise tuning of the anisotropy of the confining potential in our experiment. One crucial result in this study is our ability to achieve and manipulate orbital hybridization between distinct quasibound states by precisely adjusting the anisotropy of the confining potential in this way.

Observed in the graphene/$WSe_2$ heterostructure, two exemplary QDs are presented in Figs. 2b, 2c: a circular QD and an elliptical QD, respectively. Atomic-resolved STM measurements (top panels of Fig. 2d) along with their fast Fourier transform (FFT) images (bottom panels of Fig. 2d) clearly confirm the phases of $WSe_2$ on and off the interfacial islands as 1T' and 2H, respectively. These phases of $WSe_2$ substrates jointly generate well-defined electrostatic potentials in the supported graphene, forming quantum-confined quasibound states. The quasibound states in the circular and elliptical graphene QDs are summarized in the scanning tunneling spectroscopy (STS) maps in Figs. 2e, 2f, reflecting the real-space modulation of LDOS of the quasibound states.

For the circular graphene QD depicted in Fig. 2b, we present the first four quasibound states: s1, p1, s2, and d1 in Fig. 2e, ordered by energy increasing. In the presence of the Coulomb potential, states s1 (0, 1) and s2 (0, 2) are identified as atomic collapse states[21,30] (ACSs), induced by strong electron attraction to the QD center[21,30,43,44]. On the other hand, states p1 (1, 1) and d1 (2, 1) represent whispering gallery modes[22,30] (WGMs), where electrons follow coherent circular paths due to Klein scattering along the potential barrier, leading to their distribution along the QD edge[22,30]. These experimental results are consistent with numerical calculations and analytical wavefunctions shown in Fig. 1e and Extended data Fig. 1, validating the observed behaviors of quasibound states.

The anisotropy degree is introduced by the elliptical graphene QD in Fig. 2c. The quasibound states exhibit the same energy sequence as that of the circular QD, labelled by s1, p1, sd+, sd-. As shown in Fig. 2f, the first ACS s1 (0, 1) and first WGM p1 (1, 1) also prominently distribute in the center and in the edge, respectively, demonstrating almost no hybridization. Remarkably, the second ACS s2 (0, 2) and second WGM d1 (2, 1) are hybridized by the anisotropic potential. Due to their close energies, the evident sd hybridization is observed. Although the ACS and WGM are viewed to origin from essentially different mechanisms[21,22], they are linked by the anisotropy-induced orbital hybridization: The hybridized states each have both ACS s2 component and WGM d1 component, as described by Eq. (1). After the hybridization, the states exhibit π-period distributions: sd+ distributes in the minor axis exhibiting a $\theta$ shape, while sd- distributes in the major axis exhibiting a rotated $\theta$ shape. These experimental observations of hybridized states are well supported by the numerical calculations and analytical wavefunctions (Fig. 1f and Extended data Fig. 1). Therefore, our result explicitly demonstrates the presence of sd hybridization induced by the anisotropy of confining potential in the artificial atoms.

**Energy splitting**

The orbital hybridization between states s2 (0, 2) and d1 (2, 1) can be described by the effective Hamiltonian in the s2, d1 basis $H_{sd} = \begin{pmatrix} E_{(0,2)} & T \\ T & E_{(2,1)} \end{pmatrix}$. Here, $E_{(0,2)}$ and $E_{(2,1)}$ represent the energies of the unhybridized states s2 and d1, respectively, and $T$ denotes their hybridization coupling strength (derived in Supplementary information). The eigenenergies of hybridized states sd+, sd- are $E_{3(4)} = \frac{E_{(0,2)}+E_{(2,1)}}{2} \pm \sqrt{\left(\frac{E_{(0,2)}-E_{(2,1)}}{2}\right)^2 + T^2}$, where $E_3$ and $E_4$ respectively correspond to plus and minus signs. As anisotropy strengthens, hybridization $T$ increases, resulting in a greater energy difference $E_3 - E_4$, i.e., the hybridized states sd+, sd- progressively split in energy.

To clearly illustrate this phenomenon, we present numerically calculated LDOS maps for four different graphene QDs with increasing anisotropy in Fig. 3a. In Fig. 3b, we also provide the corresponding experimental space-energy STS $-d^3I/dV^3$ maps for four different graphene QDs for comparison, with various anisotropy degree tailored by the STM tip. The use of $-d^3I/dV^3$ maps enhances the visibility of $dI/dV$ peaks. To compare the energy splitting across different graphene QDs, we scale the anisotropy degree by $dr' = \frac{r_2-r_1}{r_2+r_1}$ (where $r_1$ is the minor radius and $r_2$ is the major radius) and make the energy scale $E' = \frac{E-(E_3+E_4)/2}{E_2-(E_3+E_4)/2}$ ($E_2$ is the energy of state p1) according to average energy spacing.

To highlight the energy splitting, we extract the energies of quasibound states and scale them to $E'$ as ordinate (the extracting method and discussion on STM tip potential[45] are provided in Supplementary information and Figs. S6, S7). By doing this, the energies $E_3', E_4'$ scaled from $E_3, E_4$ are respectively positive and negative. For the almost circular sample with $dr' = 0.04$, the rotation symmetry is almost preserved and the hybridization is almost absent. The first four states are successively s1 (0, 1), p1 (1, 1), s2 (0, 2), d1 (2, 1), consistent with previous observations[22,25,26,30]. As marked by the blue and red dots and black dashed lines in Figs. 3a, 3b, the states s2 and d1 are indeed close in energy for such a small $dr'$. As $dr'$ increases, these states hybridize into sd+, $\psi_{(0,2)} + \alpha\psi_{(2,1)}$, and sd-, $-\alpha\psi_{(0,2)} + \psi_{(2,1)}$, indeed with $E'_3, E'_4$ gradually separated (a more detailed evolution is provided in Fig. S8 in Supplementary information). For clarity, we plot the numerical LDOS-energy curves and experimental $-d^3I/dV^3$-energy curves in Figs. 3c, 3d. The state sd+ is dominated by ACS (0, 2), and is reflected by the blue curves extracted from QD center. The state sd- is dominated by WGM (2, 1), and is reflected by the red curves extracted from QD edge. The energies of the sd+ and sd- are marked by the blue and red arrows. The energy splitting versus $dr'$ increasing is further verified by both experimental measurements and numerical calculations in Figs. 3c, 3d.

We conduct extensive experiments, presenting additional experimental data and extra samples in Extended data Figs. 2-9 and Figs. S9, S10 in Supplementary information. The energy ordering and LDOS distributions of quasibound states closely resemble the results in Figs. 2e, 2f, 3b. As the signature of orbital hybridization, the hybridized states sd+ and sd- also consist of s-orbital ACS (0, 2) and d-orbital WGM (2, 1). sd+ and sd- are observed to distribute along the minor and major axes of the ellipse, respectively. For each sample, we extract and scale the energy $E'_3, E'_4$ of hybridized states sd+, sd-, respectively (the unscaled energy values are listed in Table S1 of Supplementary information). We plot $E'_3$ (positive ordinate) and $E'_4$ (negative ordinate) versus $dr'$ together in Fig. 3e (the red dots). Our experiments are performed under consistent conditions, qualitatively demonstrating the evolution of QD deformation and enhancement of orbital hybridization. Overall, the data clearly illustrate the splitting between $E'_3$ and $E'_4$ as $dr'$ increases.

On the other hand, we conduct numerical simulations on the evolution from circular QD to elliptical QD. We vary the parameter $dr'$ in equal intervals from 0 to 0.20, while keeping all the other parameters the same as those in Fig. 3a (provided in Supplementary information). For each $dr'$, we numerically calculate the space-energy LDOS maps. For five typical $dr'$, the space-energy maps and LDOS spatial distributions of quasibound states sd+ and sd- are illustrated in Extended data Fig. 10. Additionally, corresponding analytical results are included in Extended data Fig. 10 for comparison. We make the same scale as experiments and plot $E'_3$ and $E'_4$ versus $dr'$ in Fig. 3e (blue dots and blue curves), demonstrating the energy splitting from orbital hybridization that fits the experimental data well. Overall, the comprehensive research verifies the anisotropy-induced orbital hybridization and energy splitting through the mutual corroboration of experiments and numerical calculations.

**Conclusions**

In summary, we first actualize that the artificial atoms exhibit orbital hybridization by introducing the anisotropy of QD potential. Although ACSs and WGMs originate from essentially different physical mechanisms of graphene, we observe their sd hybridization due to their close energy. They recombine to form new states with LDOS distributed as a $\theta$ shape and a rotated $\theta$ shape, revealing their deep connections. Along with the QD deformation and orbital hybridization, the hybridized states gradually split in energy. Our experimental findings are bolstered by both numerical calculations and analytical derivations, collectively confirming the emergence of orbital hybridization. Our study marks a notable advancement in simulating real atomic behaviors. The findings provide a fresh perspective and underscore the potential for quantum simulation as a transformative tool in physics and engineering.

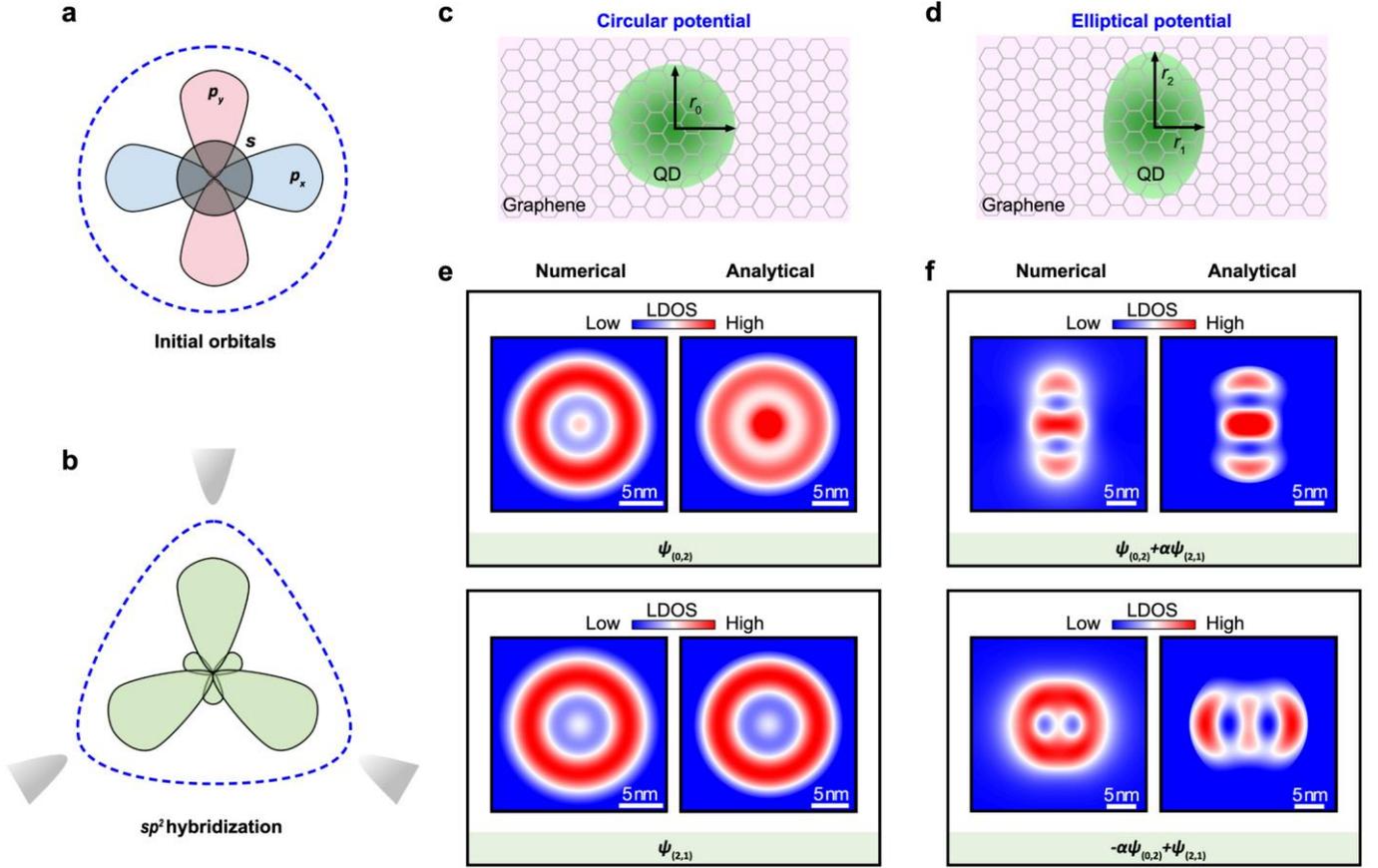

**FIG. 1. Orbital hybridization in natural atoms and artificial atoms. a.** Schematic of $s, p_x, p_y$ orbitals of natural atom under isotropic potential. **b.** Schematic of $sp^2$ hybridization induced by anisotropic potential. **c.** Schematic of confinement of graphene by a circular QD. The QD leads to a confining potential in the graphene which causes a series of quasibound states. **d.** Schematic of confinement of graphene by an elliptical QD. The anisotropy of QD potential leads to the orbital hybridization. **e.** The numerical and analytical LDOS of quasibound states s2 (0, 2) and d1 (2, 1) without hybridization. The two states are close in energy. Their wavefunctions are labeled as $\psi_{(0,2)}$ and $\psi_{(2,1)}$. **f.** The numerical and analytical LDOS of quasibound states hybridized as sd+, $\psi_{(0,2)} + \alpha\psi_{(2,1)}$, and sd-, $-\alpha\psi_{(0,2)} + \psi_{(2,1)}$. They respectively show a $\theta$ shape and a rotated $\theta$ shape.

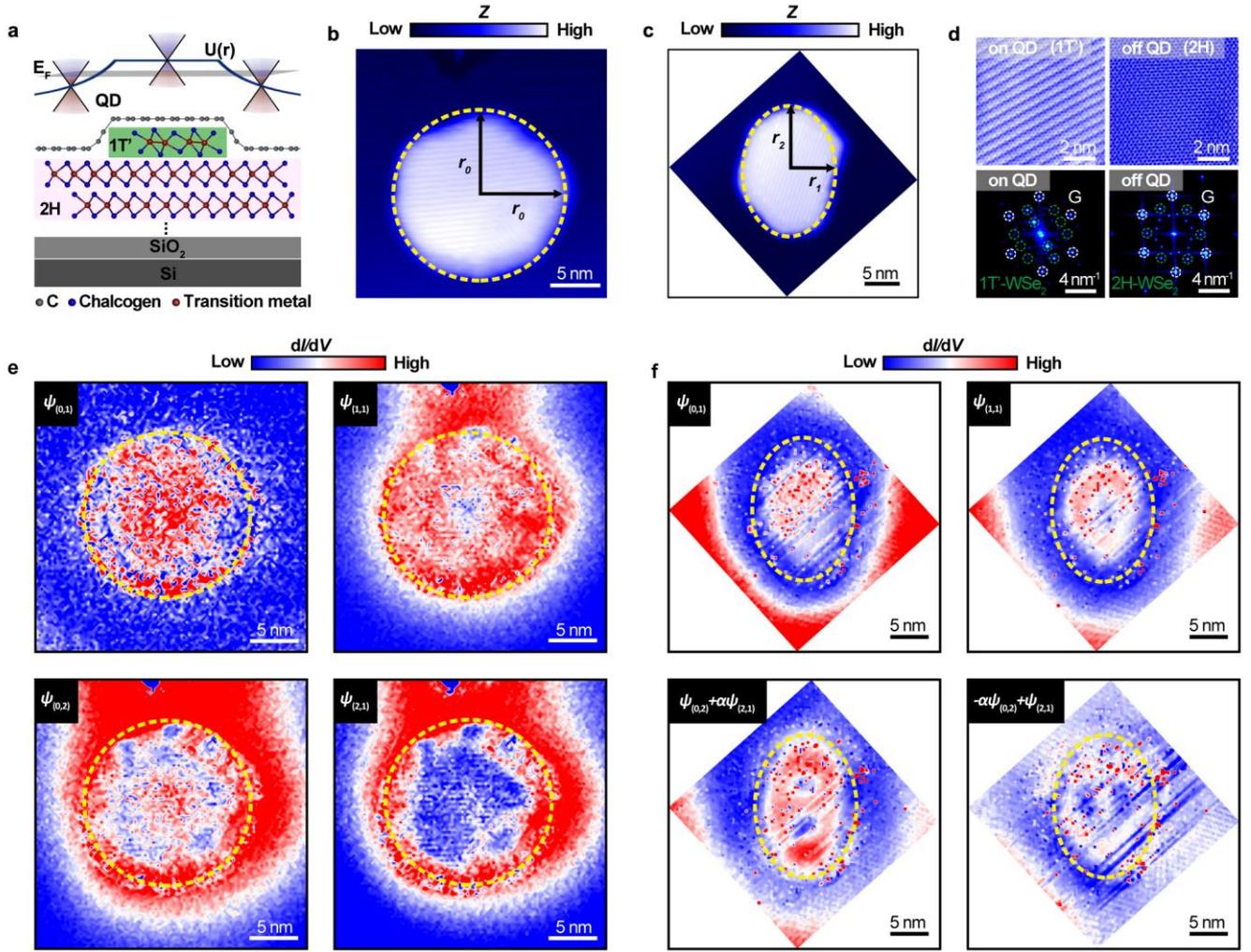

**FIG. 2. Experimental observation of hybridized orbitals. a.** Schematics of an interfacial monolayer 1T'-phase TMD island in a graphene/TMD heterostructure (bottom panel) and local electrostatic potentials in graphene induced by the interfacial 1T'-phase TMD island (top panel). **b.** A STM image ($V_b$ = 600 mV, $I$ = 100 pA) of the circular QD embedded in the graphene/WSe$_2$ heterostructure. The radius $r_0 \approx 8$ nm. **c.** A STM image ($V_b$ = 70 mV, $I$ = 100 pA) of the elliptical QD embedded in the graphene/WSe$_2$ heterostructure. The minor radius $r_1$ is approximately 8 nm, and the major radius $r_2$ is approximately 11 nm. **d.** Top panels: Atomic-resolved STM image on (1T' phase) and off (2H phase) the QD, respectively. Bottom panels: The FFT image obtained from the STM image on and off the QD, respectively. The white and green circles show reciprocal lattices of graphene and WSe$_2$, respectively. The unlabeled bright spots correspond to the reciprocal moiré superlattices and higher-order scattering. **e.** d$I$/d$V$ maps of different states (0, 1), (1, 1), (0, 2) and (2, 1) for a circular confinement. Orbital hybridization does not occur, thus all states exhibit rotational symmetry. **f.** d$I$/d$V$ maps of different states for an elliptical confinement. The anisotropy of confining potential results in orbital hybridization between the s-orbital and d-orbital states, giving rise to recombined states sd+ $\psi_{(0,2)} + \alpha\psi_{(2,1)}$ and sd- $-\alpha\psi_{(0,2)} + \psi_{(2,1)}$, which exhibit $\theta$-shaped and rotated $\theta$-shaped features, respectively. The yellow dashed lines show outlines of QDs in panels b-c and e-f.

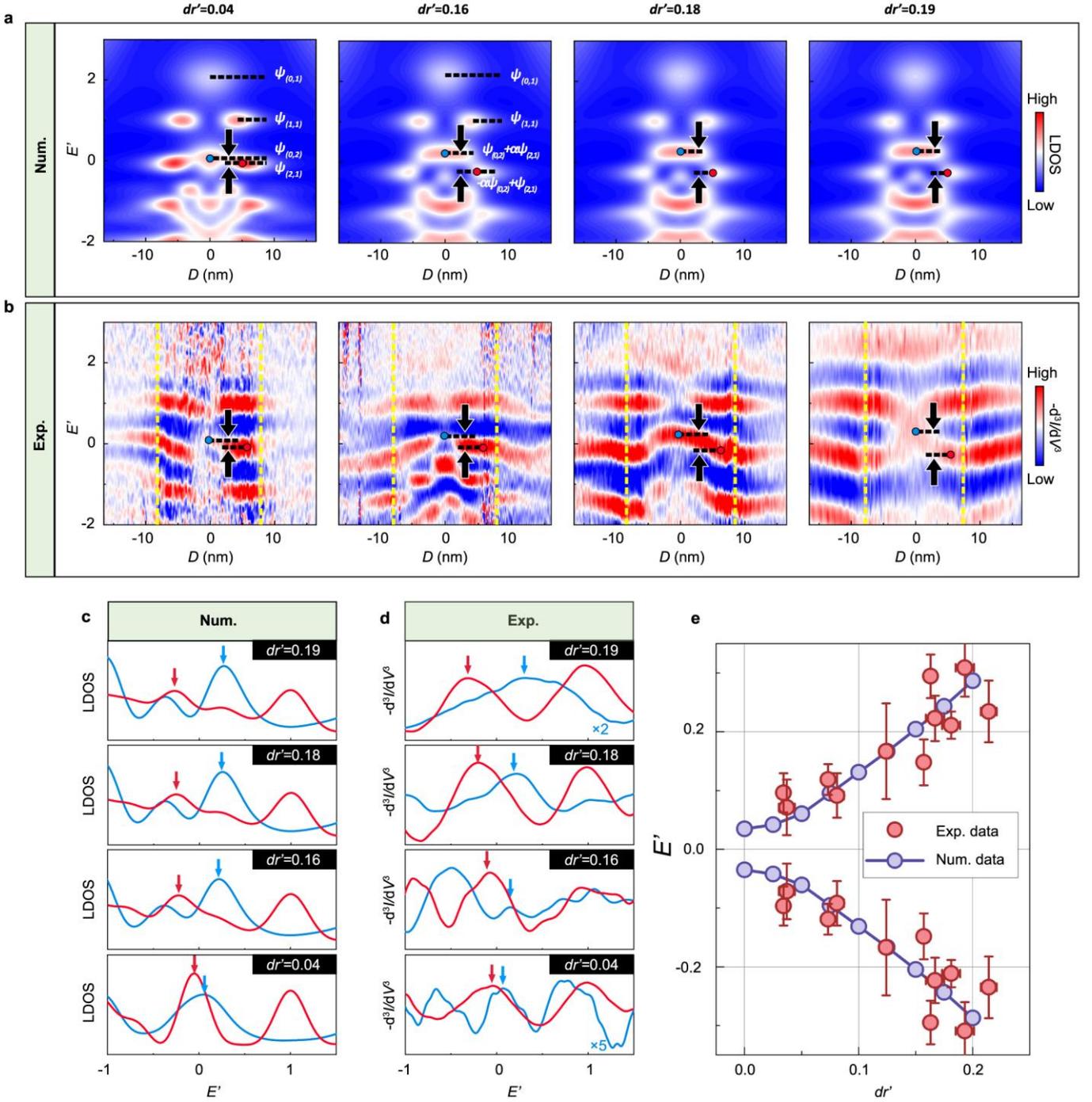

**FIG. 3. Energy splitting of hybridized states versus anisotropy degree. a and b.** The space-energy spectrums along minor axis of QDs, for 4 different anisotropy degrees $dr'$. **a** shows numerical LDOS and **b** shows experimental $-d^3I/dV^3$. Because in experiments there are differences in the size of QDs and in the Fermi velocity of graphene, dimensionless scale is made: $dr' = \frac{r_2 - r_1}{r_2 + r_1}$ represents the anisotropy degree of elliptical confinement. $E' = \frac{E - (E_3 + E_4)/2}{E_2 - (E_3 + E_4)/2}$ indicates the energy relative to the energy level spacing. In numerical results for $dr' = 0.04$ and $0.16$, the first four states are labelled by wavefunctions and black dashed lines. In all the figures, the hybridized states sd+ and sd- are respectively marked as blue and red dots. The black dashed lines and black arrows highlight their energy splitting. **c and d.** The numerical LDOS and experimental $-d^3I/dV^3$ versus $E'$. The blue and red curves are respectively extracted from the positions of blue and red dots in a and b. The LDOS and $-d^3I/dV^3$ peaks are marked by blue and red arrows. In panel d, ×2 and ×5 represent the magnification factors of the corresponding color spectra, applied for clarity. **e.** The scaled energy $E'_3, E'_4$ of sd+ and sd- versus $dr'$. The red data and blue data are respectively experimental results and numerical results. For both experimental and numerical data, the positive (negative) ordinates correspond to $E'_3$ ($E'_4$). The experimental data are extracted from b, Extended data Figs. 2-9, and Figs. S9, S10 in Supplementary information, with error bars reflecting their standard error. The numerical data are extracted from the calculations for $dr'$ from 0 to 0.20.

## Methods

### Samples Preparation

In our experiments, we obtain high-quality graphene/TMD heterostructures using a transfer technique involving graphene monolayer transfer onto mechanically-exfoliated thick 2H-phase TMD sheets. The specific method are as follows:

**CVD Growth of graphene.** Large-area graphene monolayer films are synthesized on a 20 × 20 mm$^2$ polycrystalline copper (Cu) foil (Alfa Aesar, 25 μm thick) using a low-pressure chemical vapor deposition (LPCVD) method. The Cu foil is heated from room temperature to 1030°C within 30 minutes and annealed at 1030°C for 12 hours, with a constant flow of 50 sccm (Standard Cubic Centimeter per Minutes) of argon (Ar) and 50 sccm of hydrogen (H$_2$) maintained throughout the process. Then, 5 sccm of methane (CH$_4$) is introduced and maintained for 20 minutes to facilitate the growth of high-quality large-area graphene monolayer. Finally, the sample is naturally cooled down to room temperature.

**Construction of graphene/TMD heterostructure.** The as-grown graphene monolayer is transferred onto mechanically-exfoliated thick TMD (WSe$_2$ or MoSe$_2$) substrates using the polymethyl methacrylate (PMMA)-assisted method. Firstly, PMMA is uniformly coated onto the Cu foil with graphene monolayer. Then, the sample is placed on the surface of an ammonium persulfate solution to etch away the Cu layer at the bottom of the sample, thereby separating the PMMA/graphene film from the Cu foil. Subsequently, the PMMA/graphene film is rinsed with deionized water for several hours. The bulk TMDs (Shanghai Onway Technology Co., Ltd) are mechanically exfoliated into thick TMD sheets and transferred onto SiO$_2$/Si wafer. The cleaned PMMA/graphene is placed on the SiO$_2$/Si wafer pre-stacked with TMD sheets, left to air dry, and then acetone is used to remove the PMMA.

**Preparation of graphene/TMD quantum dots.** After obtaining high-quality graphene/TMD (2H phase) heterostructures, nanoscale monolayer TMD islands are created at the heterostructure interface by using scanning tunneling microscope (STM) tip pulses. During the formation process, the TMD islands undergo a structural phase transition to the 1T' phase, resulting in different doping effects of the TMD islands and the surrounding 2H-phase TMD on graphene, thereby creating well-defined nanoscale graphene quantum dots.

### STM and STS Measurements

STM/STS measurements are performed in low-temperature (77 K and 4.2 K) and ultrahigh-vacuum (~10$^{-10}$ Torr) scanning probe microscopes [USM-1500 (77 K) and USM-1300 (4.2 K)] from UNISOKU. The tips are obtained by chemically etching tungsten wires to minimize tip-induced band bending effects of graphene. The differential conductance (d$I$/d$V$) measurements are taken by a standard lock-in technique with an ac bias modulation of 5 mV and 793 Hz signal added to the tunneling bias.

### Data availability

Source data are provided with this paper.

### Code availability

Codes are provided in the Supplementary information.

### Acknowledgments

This work was financially supported by the National Key R and D Program of China (Grant Nos. 2021YFA1400100 and 2021YFA1401900), the National Natural Science Foundation of China (Grants Nos. 12374034, 11921005, 12141401, 12425405 and 12404198), the Innovation Program for Quantum Science and Technology (Grant No. 2021ZD0302403), "the Fundamental Research Funds for the Central Universities" (Grant No. 310400209521), the China National Postdoctoral Program for Innovative Talents (Grant No. BX20240040) and the China Postdoctoral Science Foundation (2023M740296). The computational resources are supported by High-performance Computing Platform of Peking University. The devices were fabricated using the transfer platform from Shanghai Onway Technology Co., Ltd.

**Author contributions**

L.H. and Q.S. conceived the work and designed the research strategy. Y.M. carried out the analytical analysis and numerical calculations under the supervision of Q.S. H.-Y.R., X.-F.Z. and Y.-N.R. fabricated the samples and performed the measurements. H.-Y.R., Y.-N.R. and L.H. analyzed the experimental data. All authors wrote the paper together.

**Competing interests**

The authors declare no competing interests.

**Additional information**

**Supplementary information** The online version contains supplementary material available.
**Correspondence and requests for materials** should be addressed to Ya-Ning Ren, Lin He, or Qing-Feng Sun.

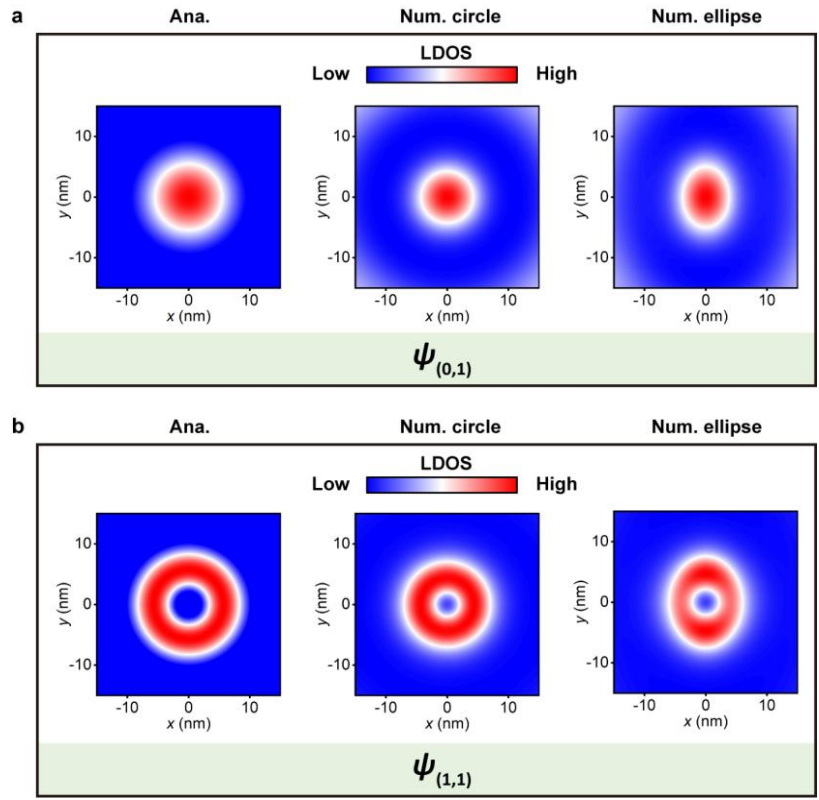

**Extended data Fig. 1. Analytical and numerical LDOS of quasibound states (0, 1) and (1, 1). a and b.** The distribution of states (0, 1) and (1, 1), respectively. Left panel: The LDOS maps from the analytically solved wavefunctions. Middle panel: The numerically calculated LDOS maps for a circular QD. Right panel: The numerically calculated LDOS maps for an elliptical QD.

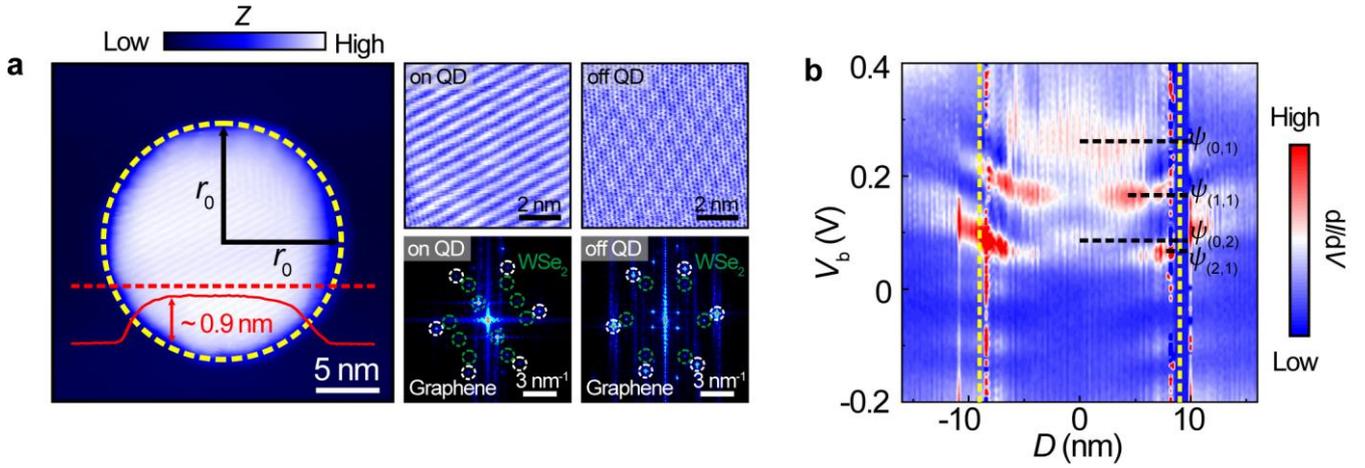

**Extended data Fig. 2. Almost unhybridized orbital states under an almost circular confinement with $dr' = 0.03$. a.** Left panel: A STM image ($V_b$ = 600 mV, $I$ = 130 pA) of the almost circular QD with anisotropy degree $dr' = 0.03$ embedded in the graphene/WSe$_2$ heterostructure. The radius $r_0 \approx 9$ nm. The yellow dashed line shows outline of the circular QD. The height profile along the red dashed line is shown with solid red line. Top right panels: Atomic-resolved STM image on (1T' phase) and off (2H phase) the QD, respectively. Bottom right panels: The FFT image obtained from the STM image on and off the QD, respectively. The white and green circles show reciprocal lattices of graphene and WSe$_2$, respectively. The unlabeled bright spots correspond to the reciprocal moiré superlattices and higher-order scattering. **b.** The d$I$/d$V$ spectroscopic map versus the spatial position along axis of the almost circular QD. Orbital states can be clearly observed in the QD. The first four states are labelled by wavefunctions and black dashed lines. The two yellow dashed lines mark the size of the QD.

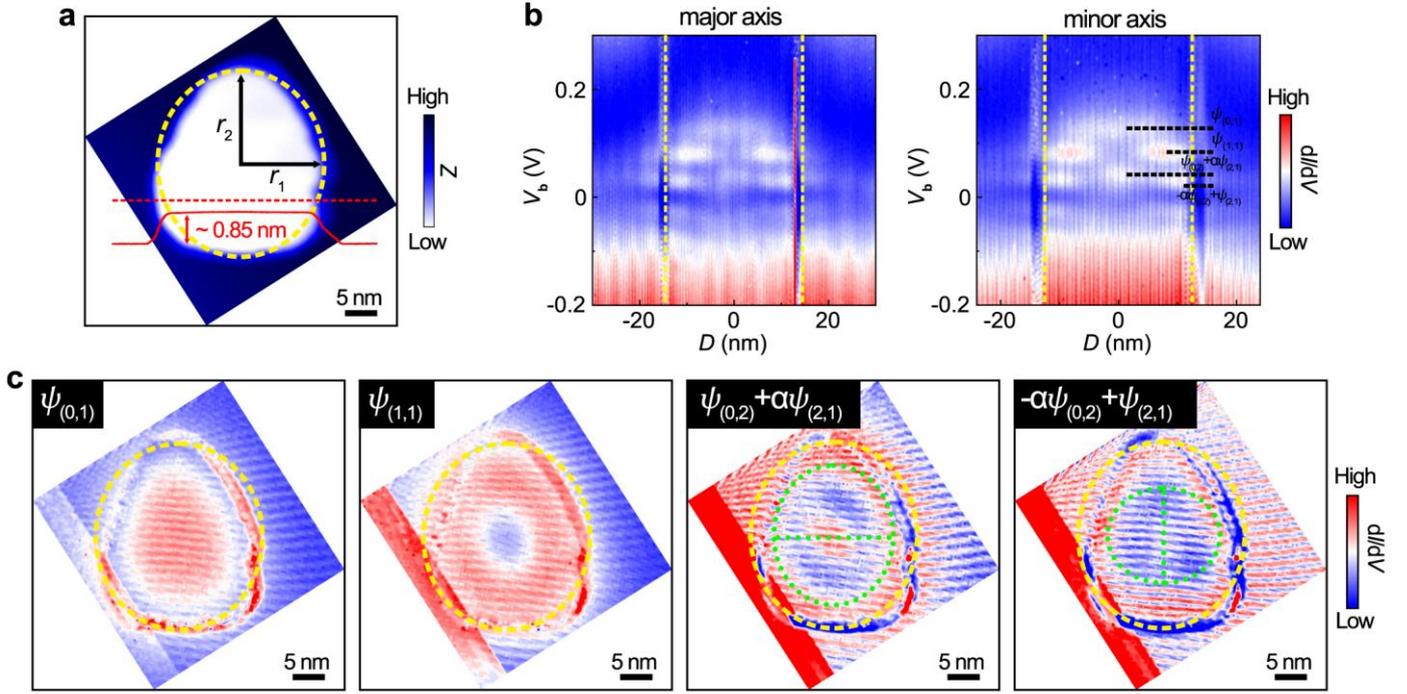

**Extended data Fig. 3. Hybridized orbital states under an elliptical confinement with $dr' = 0.07$. a.** A STM image ($V_b$ = 600 mV, $I$ =100 pA) of the elliptical QD with anisotropy degree $dr' = 0.07$ embedded in the graphene/WSe$_2$ heterostructure. The minor radius $r_1$ is approximately 12.5 nm, and the major radius $r_2$ is approximately 14.5 nm. The height profile along the red dashed line is shown with solid red line. **b.** The d$I$/d$V$ spectroscopic maps versus the spatial position along the major axis (left panel) and minor axis (right panel) of the elliptical QD, respectively. Orbital states can be clearly observed in the QD. The first four states are labelled by wavefunctions and black dashed lines in the right panel. The yellow dashed lines mark the size of the QD. **c.** d$I$/d$V$ maps of different orbital states. For an elliptical confinement, the anisotropy of confining potential results in orbital hybridization between the s-orbital and d-orbital states, giving rise to new states sd+ $\psi_{(0,2)} + \alpha\psi_{(2,1)}$ and sd- $-\alpha\psi_{(0,2)} + \psi_{(2,1)}$, which exhibit $\theta$-shaped and rotated $\theta$-shaped features marked by the green dotted lines, respectively. The yellow dashed lines show outlines of the elliptical QD.

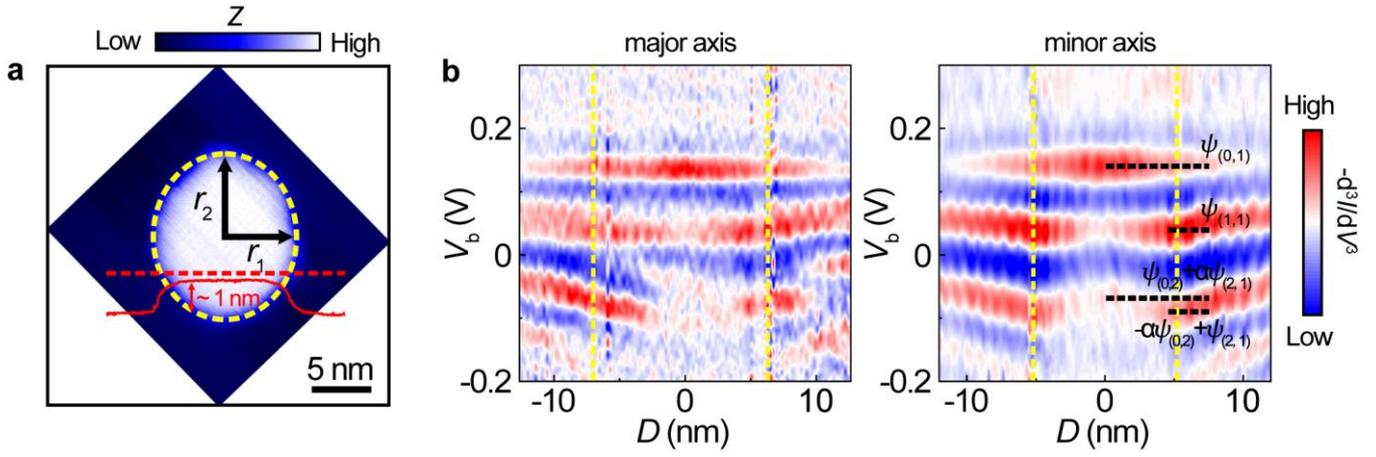

**Extended data Fig. 4. Hybridized orbital states under an elliptical confinement with $dr' = 0.08$. a.** A STM image ($V_b$ = 450 mV, $I$ = 300 pA) of the elliptical QD with anisotropy degree $dr' = 0.08$ embedded in the graphene/WSe$_2$ heterostructure. The minor radius $r_1$ is approximately 6 nm, and the major radius $r_2$ is approximately 7 nm. The height profile along the red dashed line is shown with solid red line. **b.** The $-d^3I/dV^3$ spectroscopic maps versus the spatial position along the major axis (left panel) and minor axis (right panel) of the elliptical QD, respectively. Orbital states can be clearly observed in the QD. The first four states are labelled by wavefunctions and black dashed lines in the right panel. The yellow dashed lines mark the size of the QD.

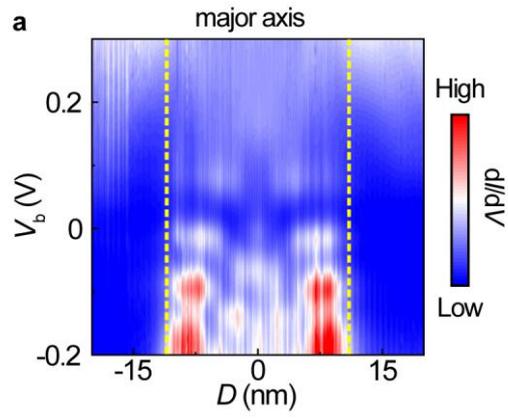

**Extended data Fig. 5. Hybridized orbital states under an elliptical confinement with $dr' = 0.16$.** **a.** The d$I$/d$V$ spectroscopic map versus the spatial position along the major axis of the elliptical QD in Fig. 2c. Orbital states can be clearly observed in the QD. The two yellow dashed lines mark the size of the QD.

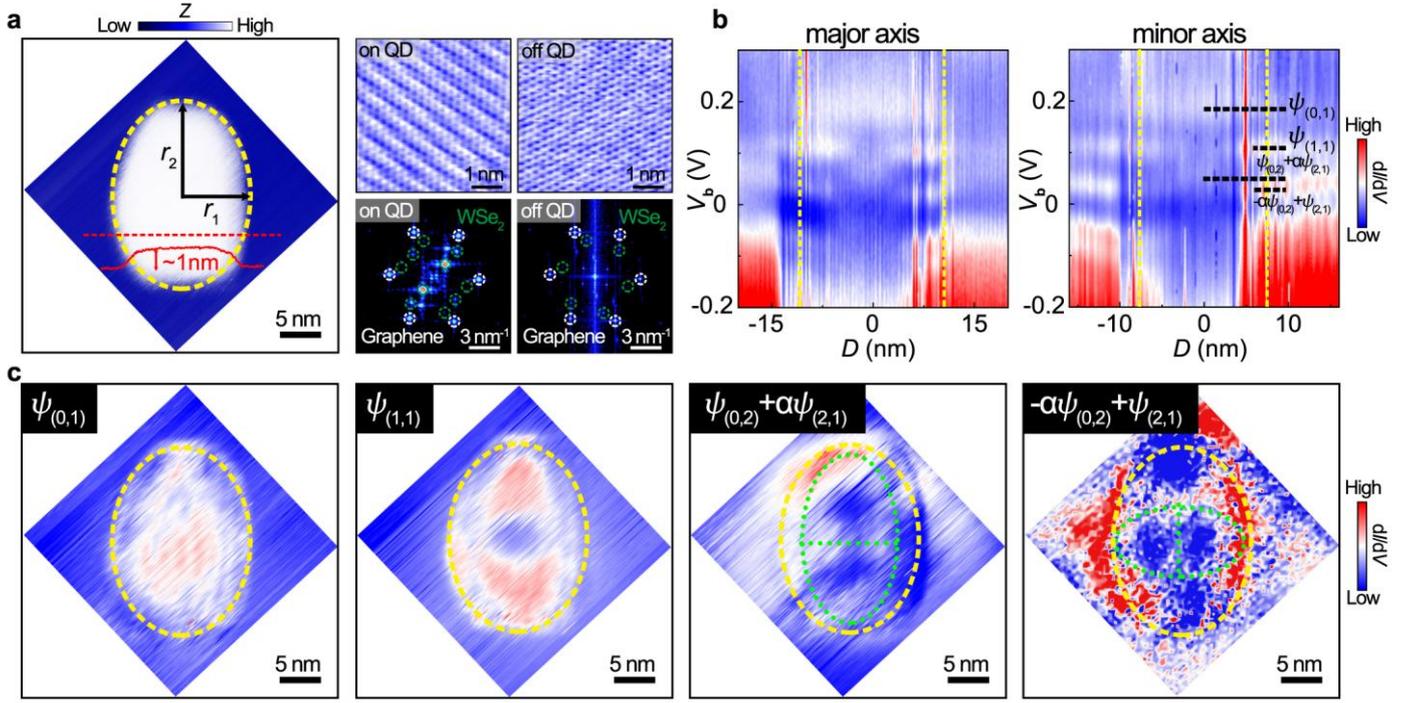

**Extended data Fig. 6. Hybridized orbital states under an elliptical confinement with $dr' = 0.17$. a.** Left panel: A STM image ($V_b$ = 500 mV, $I$ = 100 pA) of the elliptical QD with anisotropy degree $dr' = 0.17$ embedded in the graphene/WSe$_2$ heterostructure. The minor radius $r_1$ is approximately 7.6 nm, and the major radius $r_2$ is approximately 10.7 nm. The height profile along the red dashed line is shown with solid red line. Top right panels: Atomic-resolved STM image on (1T' phase) and off (2H phase) the QD, respectively. Bottom right panels: The FFT image obtained from the STM image on and off the QD, respectively. The white and green circles show reciprocal lattices of graphene and WSe$_2$, respectively. The unlabeled bright spots correspond to the reciprocal moiré superlattices and higher-order scattering. **b.** The d$I$/d$V$ spectroscopic maps versus the spatial position along the major axis (left panel) and minor axis (right panel) of the elliptical QD, respectively. Orbital states can be clearly observed in the QD. The first four states are labelled by wavefunctions and black dashed lines in the right panel. The yellow dashed lines mark the size of the QD. **c.** d$I$/d$V$ maps of different orbital states. For an elliptical confinement, the anisotropy of confining potential results in orbital hybridization between the s-orbital and d-orbital states, giving rise to new states sd+ $\psi_{(0,2)} + \alpha\psi_{(2,1)}$ and sd- $-\alpha\psi_{(0,2)} + \psi_{(2,1)}$, which exhibit $\theta$-shaped and rotated $\theta$-shaped features marked by the green dotted lines, respectively. The yellow dashed lines show outlines of the elliptical QD.

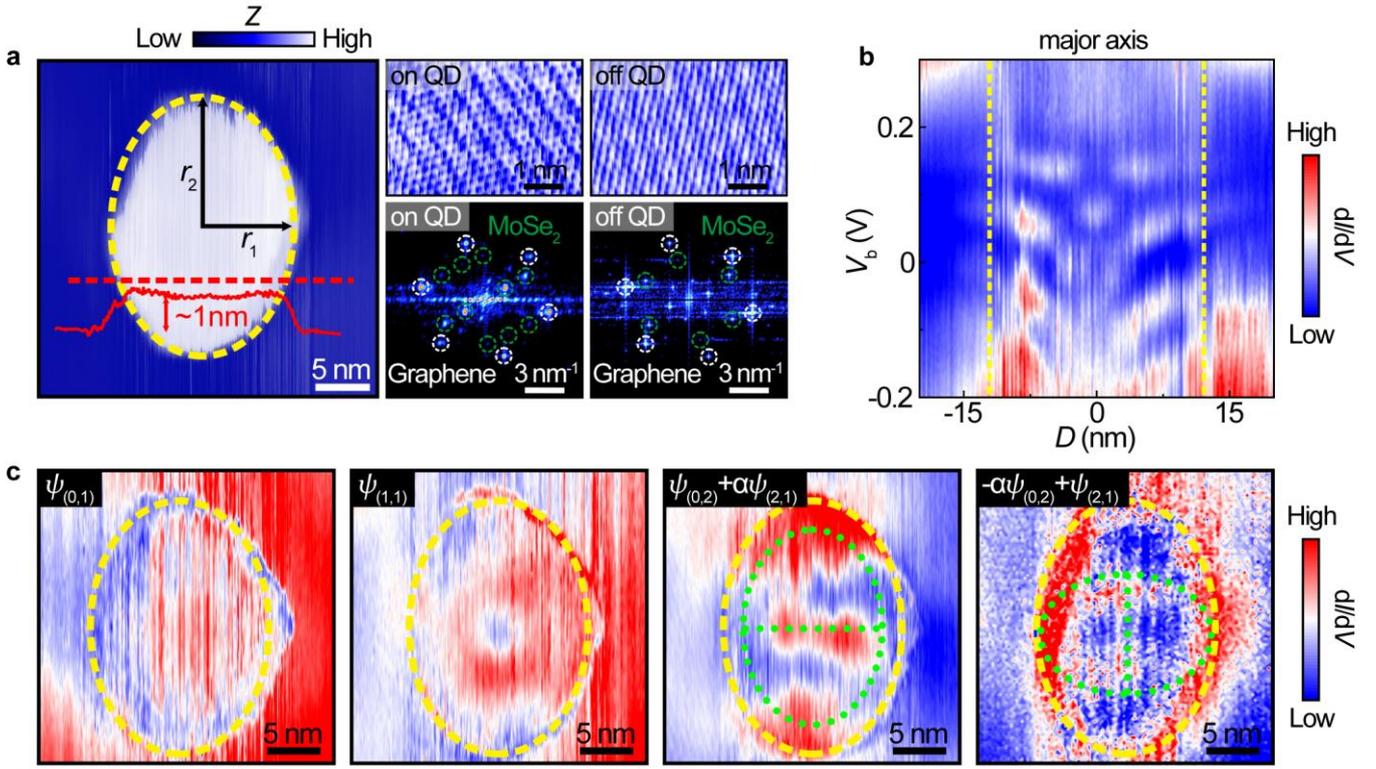

**Extended data Fig. 7. Hybridized orbital states under an elliptical confinement with $dr' = 0.18$. a.** Left panel: A STM image ($V_b$ = 400 mV, $I$ = 120 pA) of the elliptical QD with anisotropy degree $dr' = 0.18$ embedded in the graphene/MoSe$_2$ heterostructure. The minor radius $r_1$ is approximately 8.4 nm, and the major radius $r_2$ is approximately 12.1 nm. The height profile along the red dashed line is shown with solid red line. Top right panels: Atomic-resolved STM image on (1T' phase) and off (2H phase) the QD, respectively. Bottom right panels: The FFT image obtained from the STM image on and off the QD, respectively. The white and green circles show reciprocal lattices of graphene and MoSe$_2$, respectively. The unlabeled bright spots correspond to the reciprocal moiré superlattices and higher-order scattering. **b.** The d$I$/d$V$ spectroscopic map versus the spatial position along the major axis of the elliptical QD. Orbital states can be clearly observed in the QD. The two yellow dashed lines mark the size of the QD. **c.** d$I$/d$V$ maps of different orbital states. For an elliptical confinement, the anisotropy of confining potential results in orbital hybridization between the s-orbital and d-orbital states, giving rise to new states sd+ $\psi_{(0,2)} + \alpha\psi_{(2,1)}$ and sd- $-\alpha\psi_{(0,2)} + \psi_{(2,1)}$, which exhibit $\theta$-shaped and rotated $\theta$-shaped features marked by the green dotted lines, respectively. The yellow dashed lines show outlines of the elliptical QD.

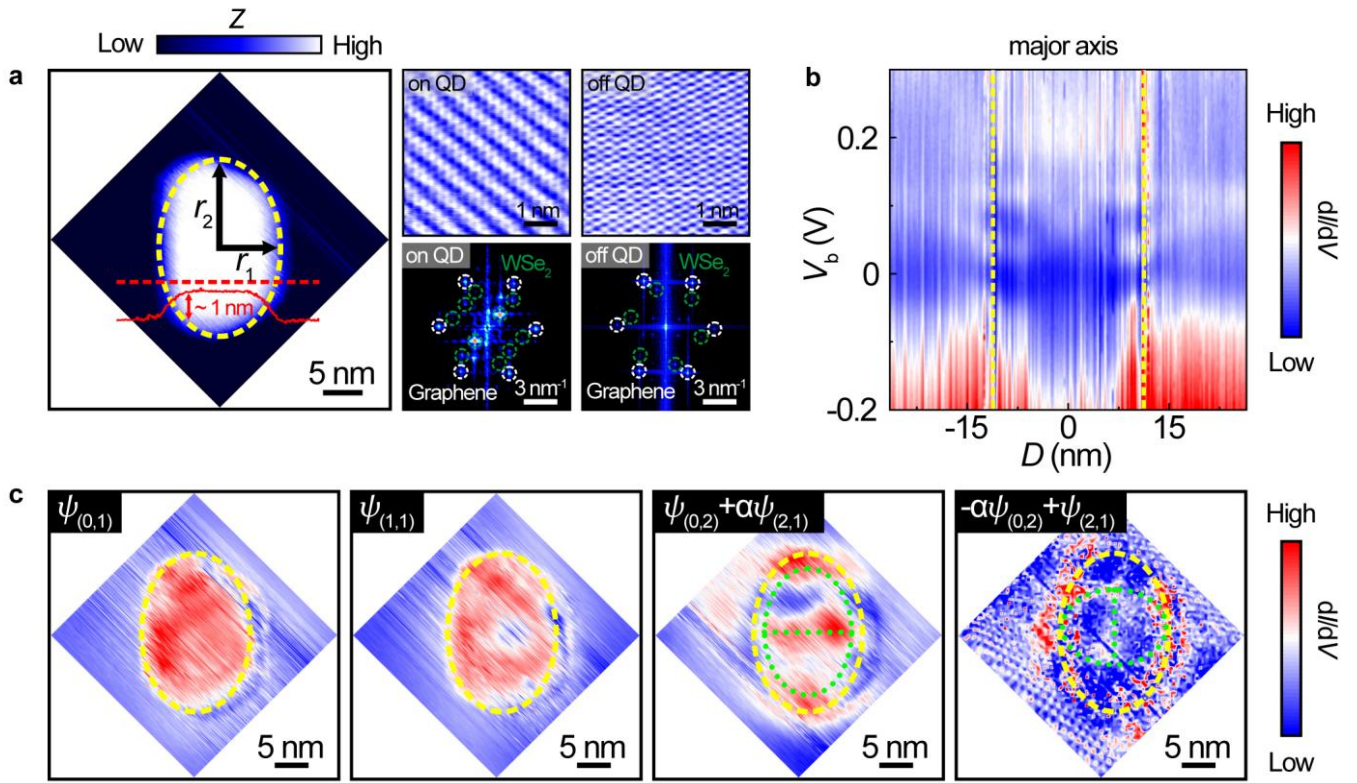

**Extended data Fig. 8. Hybridized orbital states under an elliptical confinement with $dr' = 0.19$. a.** Left panel: A STM image ($V_b$ = 500 mV, $I$ = 100 pA) of the elliptical QD with anisotropy degree $dr' = 0.19$ embedded in the graphene/WSe$_2$ heterostructure. The minor radius $r_1$ is approximately 7.6 nm, and the major radius $r_2$ is approximately 11.2 nm. The height profile along the red dashed line is shown with solid red line. Top right panels: Atomic-resolved STM image on (1T' phase) and off (2H phase) the QD, respectively. Bottom right panels: The FFT image obtained from the STM image on and off the QD, respectively. The white and green circles show reciprocal lattices of graphene and WSe$_2$, respectively. The unlabeled bright spots correspond to the reciprocal moiré superlattices and higher-order scattering. **b.** The d$I$/d$V$ spectroscopic map versus the spatial position along the major axis of the elliptical QD. Orbital states can be clearly observed in the QD. The two yellow dashed lines mark the size of the QD. **c.** d$I$/d$V$ maps of different orbital states. For an elliptical confinement, the anisotropy of confining potential results in orbital hybridization between the s-orbital and d-orbital states, giving rise to new states sd+ $\psi_{(0,2)} + \alpha\psi_{(2,1)}$ and sd- $-\alpha\psi_{(0,2)} + \psi_{(2,1)}$, which exhibit $\theta$-shaped and rotated $\theta$-shaped features marked by the green dotted lines, respectively. The yellow dashed lines show outlines of the elliptical QD.

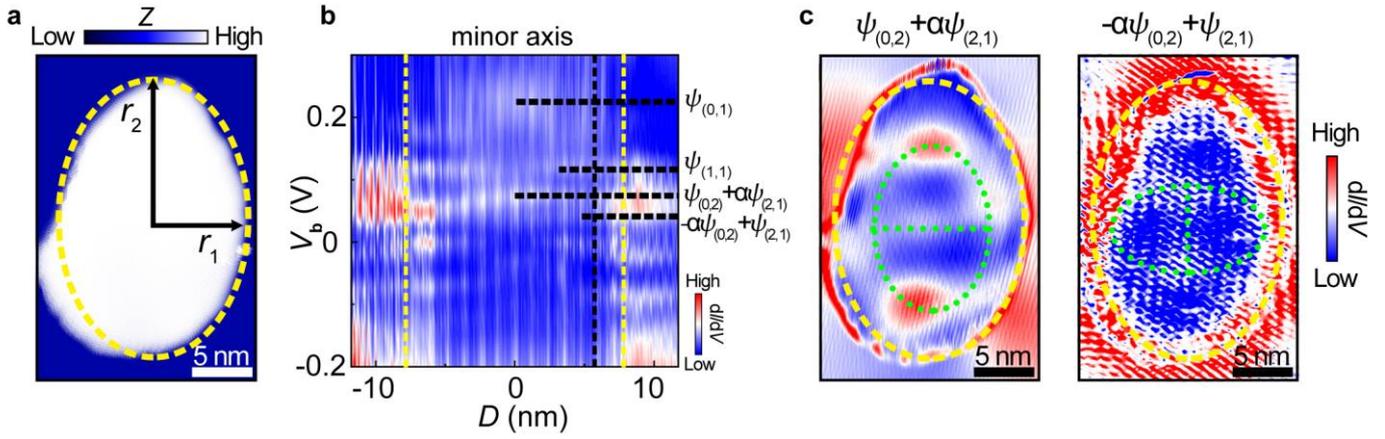

**Extended data Fig. 9. Hybridized orbital states under an elliptical confinement with $dr' = 0.21$. a.** A STM image ($V_b$ = -140 mV, $I$ = 150 pA) of the elliptical QD with anisotropy degree $dr' = 0.21$ embedded in the graphene/WSe$_2$ heterostructure. The minor radius $r_1$ is approximately 7.8 nm, and the major radius $r_2$ is approximately 12 nm. **b.** The d$I$/d$V$ spectroscopic map versus the spatial position along the minor axis of the elliptical QD. Orbital states can be clearly observed in the QD. The first four states are labelled by wavefunctions and black dashed lines. The two yellow dashed lines mark the size of the QD. **c.** d$I$/d$V$ maps of different orbital states. For an elliptical confinement, the anisotropy of confining potential results in orbital hybridization between the s-orbital and d-orbital states, giving rise to new states sd+ $\psi_{(0,2)} + \alpha\psi_{(2,1)}$ and sd- $-\alpha\psi_{(0,2)} + \psi_{(2,1)}$, which exhibit $\theta$-shaped and rotated $\theta$-shaped features marked by the green dotted lines, respectively. The yellow dashed lines show outlines of the elliptical QD.

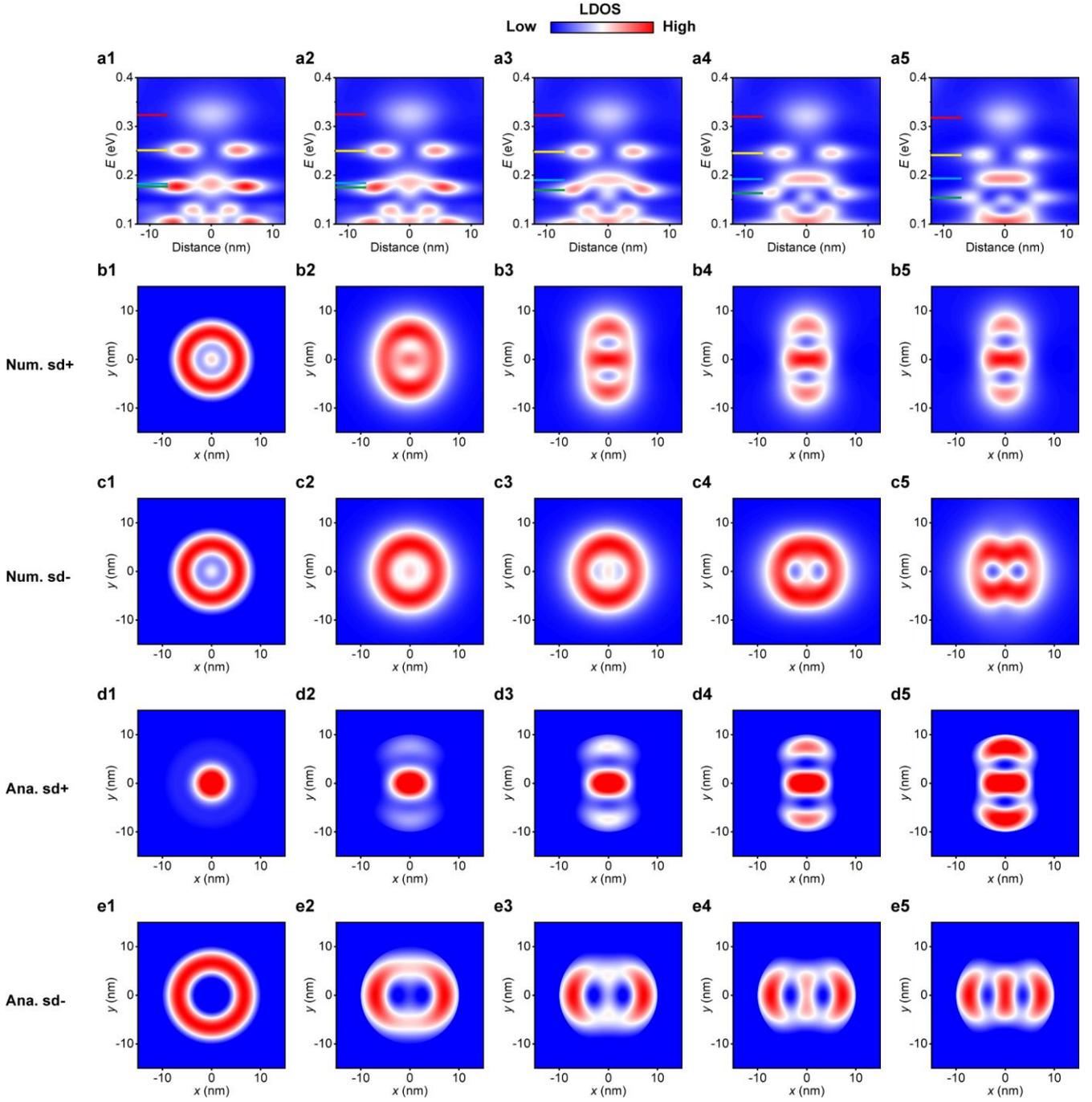

**Extended data Fig. 10. LDOS evolution of hybridized states from s2 (0, 2) and d1 (2, 1).** From the leftmost panel to the rightmost panel: $dr' = 0, 0.05, 0.10, 0.15, 0.20$, the other parameters are the same and shown in Supplementary information. **a1-a5.** The space-energy LDOS for the five potentials, along the minor axis of ellipse. The red, yellow, green, and blue color bars respectively indicate the energy of quasibound states s1, p1, sd+, sd-. **b1-b5, c1-c5.** The numerically calculated spatial distributions of hybridized states sd+, sd- evolved as the QD is deformed from circle to ellipse. **d1-d5, e1-e5.** The spatial distributions of sd+ and sd- obtained from the combination of analytical wavefunctions $\psi_{sd+} = \psi_{(0,2)} + \alpha\psi_{(2,1)}$ and $\psi_{sd-} = -\alpha\psi_{(0,2)} + \psi_{(2,1)}$. From left to right: $\alpha = 0.0, 0.3, 0.5, 0.7, 0.9$.

## Supplementary Notes

### I. Analytical solution for confined states.

Let us first just consider the graphene without any electrostatic potential. It can be described by the Hamiltonian[1]:

$$H = \hbar v_F \begin{pmatrix} 0 & k_x - ik_y \\ k_x + ik_y & 0 \end{pmatrix} = \hbar v_F \begin{pmatrix} 0 & -ie^{-i\phi}\left(\frac{\partial}{\partial r} - \frac{i}{r}\frac{\partial}{\partial \phi}\right) \\ -ie^{i\phi}\left(\frac{\partial}{\partial r} + \frac{i}{r}\frac{\partial}{\partial \phi}\right) & 0 \end{pmatrix}, \quad (S1)$$

where $v_F$ is the Fermi velocity, $\hbar$ is the reduced Planck constant, $k_{x(y)} = -i\frac{\partial}{\partial x(y)}$ is the wave vector operator. It is further transformed to polar coordinate system, with $r$ the radial coordinate and $\phi$ the polar angle. The wavefunction with orbital quantum number $m$ is $\psi_m = \begin{pmatrix} a_m(r)e^{im\phi} \\ ib_m(r)e^{i(m+1)\phi} \end{pmatrix}$.[2,3] Here the basis of $\psi_m$ is sublattices A and B. $a_m(r)$ and $b_m(r)$ are the radial components of A and B, respectively. Solving the Schrödinger equation $H\psi_m = E\psi_m$, one can find that $a_m(r)$ and $b_m(r)$ satisfy Bessel equations[2]. The solution is

$$\psi_m = \begin{pmatrix} j_m(|E/\hbar v_F|r)e^{im\phi} \\ i\,sgn(E)\,j_{m+1}(|E/\hbar v_F|r)e^{i(m+1)\phi} \end{pmatrix}. \quad (S2)$$

Here $j_m$ is the $m$-order Bessel function of the first kind, and $E$ is the energy.

When there is a circular confining potential with energy height $V_0$ and radius $R$, for exact solution, one should set a series of boundary conditions which lead to complicated operations and formulas[2]. But for a concise physical picture, we here approximately set a boundary condition that the wavefunction of A sublattice is zero at $r = R$. This boundary condition leads to $j_m(|(E-V_0)/\hbar v_F|R) = 0$. Note that in Eq. (S2), $E$ is replaced by $E - V_0$ by the potential. The boundary condition is satisfied for only discrete energy levels: The $n$-th quasibound state for orbital number $m$ is derived to be $|E_{m,n} - V_0| = \frac{\hbar v_F x_{m,n}}{R}$, where $x_{m,n}$ is the $n$-th zero point of $m$-order Bessel function. Hereafter the confining potential we study is high in the center and low in the edge. That is, we consider the hole state. Therefore, the energy $E$ of quasibound state is lower than the potential $V_0$ (the Dirac point) in the center, $sgn(E - V_0) = -1$ and $E_{m,n} = V_0 - \frac{\hbar v_F x_{m,n}}{R}$. In fact, $sgn(E - V_0) = 1$ or $-1$ does not affect the local density of states (LDOS) results. In this way, inside the confining region $r \le R$, the wavefunction of the quasibound state (m, n) is

$$\psi_{(m,n)} = \begin{pmatrix} j_m\left(\frac{x_{m,n}}{R}r\right)e^{im\phi} \\ -ij_{m+1}\left(\frac{x_{m,n}}{R}r\right)e^{i(m+1)\phi} \end{pmatrix}. \quad (S3)$$

Among the zero points of integer order Bessel functions, the smallest zero points are $x_{0,1} < x_{1,1} < x_{0,2} \approx x_{2,1}$. This means $E_{0,1} > E_{1,1} > E_{0,2} \approx E_{2,1}$, which is consistent with previous experimental and numerical studies[3–6] and our present experimental and numerical results: In order of energy from high to low, the states are (0, 1), (1, 1), (0, 2), (2, 1), indicated by s1, p1, s2, d1, respectively. s2 and d1 are close in energy. From Eq.

(S3), the analytical LDOS (i.e. the absolute square of wavefunction) of s1, p1 are plotted in Extended data Fig. 1, and the analytical LDOS of unhybridized s2, d1 are plotted in Extended data Figs. 10d1, 10e1. In Fig. 1e, the states s2 (0, 2) and d1 (2, 1) are close in energy and their energy broadening leads to their overlap. To better reflect the actual situation, we plot the $0.4|\psi_{(0,2)}|^2 + 0.6|\psi_{(2,1)}|^2$ and $0.2|\psi_{(0,2)}|^2 + 0.8|\psi_{(2,1)}|^2$ in right panels of Fig. 1e. Note that this is the direct summation of unhybridized LDOS due to energy broadening, and is different from the hybridized state like $|\sqrt{0.4}\psi_{(0,2)} + \sqrt{0.6}\psi_{(2,1)}|^2$.

## II. Orbital hybridization induced by anisotropic potential.

In the system without rotation symmetry, the potential has an extra anisotropic term $V(r,\phi)$. Thus, the quasibound states are no longer independent in angular momentum and can be coupled by $V(r,\phi)$. From the perturbation theory[7], the effective coupling between quasibound states $\psi_{(m,n)}$ and $\psi_{(l,k)}$ can be written as

$$T = \langle \psi_{(m,n)}|V(r,\phi)|\psi_{(l,k)}\rangle = \int r dr \int d\phi V(r,\phi) e^{i(l-m)\phi} [a_{m,n}^*(r) a_{l,k}(r) + b_{m,n}^*(r) b_{l,k}(r)]. \quad (S4)$$

To obtain a nonzero coupling, the integral in the polar angle space $\int d\phi V(r,\phi) e^{i(l-m)\phi}$ must be nonzero. According to the orthogonal relationship, the extra potential must contain an $e^{-i(l-m)\phi}$ term (or equivalently $\sin(l-m)\phi$, $\cos(l-m)\phi$ terms), so that the quasibound states with orbital numbers $m$ and $l$ can couple. The coupling causes the hybridization between states $\psi_{(m,n)}$ and $\psi_{(l,k)}$, which can be remarkable when the two original states are close in energy.

Note that the first two states s1 (0, 1) and p1 (1, 1) are far away from other states in energy, but the next two states s2 (0, 2) and d1 (2, 1) have very close energy and can be hybridized remarkably. Because their orbital number difference is 2, the hybridization can be induced by a $e^{-i2\phi}$ potential term, which directly corresponds to the experimental elliptical potential.

We next discuss the hybridization features between the state s2 (0, 2) and the state d1 (2, 1). Their wavefunctions are expressed as

$$\psi_{(0,2)} = \begin{pmatrix} j_0\left(\frac{x_{0,2}}{R}r\right) \\ -ij_1\left(\frac{x_{0,2}}{R}r\right)e^{i\phi} \end{pmatrix}, \quad \psi_{(2,1)} = \begin{pmatrix} j_2\left(\frac{x_{2,1}}{R}r\right)e^{i2\phi} \\ -ij_3\left(\frac{x_{2,1}}{R}r\right)e^{i3\phi} \end{pmatrix}. \quad (S5)$$

When the extra anisotropic potential (e.g. elliptical potential) is introduced, the hybridization between the s2 and d1 states occurs, leading to their recombination that forms new states sd+ and sd-:

$$\psi_{sd+} = \psi_{(0,2)} + \alpha \psi_{(2,1)}, \quad \psi_{sd-} = -\alpha \psi_{(0,2)} + \psi_{(2,1)}, \quad (S6)$$

where $\alpha \in [0,1]$ is the recombining coefficient indicating the hybridization strength. The hybridized states sd+ and sd- are the linear combination of s2 (0, 2) and d1 (2, 1). $\alpha = 0$ corresponds to the unhybridized case under a circular confinement, and $\alpha$ is increased by the deformation of confining potential from circular to elliptical. In Extended data Fig. 10d2-10d5, 10e2-10e5, for different degrees of deformation $dr' = 0.05, 0.10, 0.15, 0.20$, we use $\alpha = 0.3, 0.5, 0.7, 0.9$ to simulate the hybridization strength. The analytical result in Fig. 1f corresponds to $\alpha = 0.7$. In all the analytical results, we set $R = 10$ nm.

### III. More realistic analytical solutions.

Above we use a concise analytical method to obtain the hybridization properties. It relies on that the wavefunction for A sublattice is zero at the confining boundary. Although this may seem "crude", its results are consistent with experimental results and numerical results, thus showing the robustness of the theoretical side of the work. To a certain extent, it tends to broaden our appeal.

Here, we also use the more realistic analytical model, just the same as the previous theoretical work[2]. Via this model, we first obtain the unhybridized quasibound states under the circular confinement. Then the states are recombined (hybridized) by the elliptical perturbation $\psi_{sd+} = \psi_{(0,2)} + \alpha\psi_{(2,1)}$, $\psi_{sd-} = -\alpha\psi_{(0,2)} + \psi_{(2,1)}$. We plot these patterns and find their great consistency with the results from the concise (or "crude") analytical model, the numerical calculations, and the experiments.

For the circular confinement, the wavefunctions can be distinguished by the angular momentum (the orbital number m) $\psi_m = \begin{pmatrix} a_m(r)e^{im\phi} \\ ib_m(r)e^{i(m+1)\phi} \end{pmatrix}$. We consider the confining potential

$$V(r) = \begin{cases} V_{in}, & \text{if } r \leq R \\ V_{out}, & \text{if } r > R \end{cases}.$$

By solving the Schrödinger equation $H\psi_m = E\psi_m$, the wavefunctions are obtained. Inside the confining radius (r⩽R), the wavefunction is

$$a_m(r) = Fj_m(\kappa_{in}r),$$
$$b_m(r) = sgn(E - V_{in})Fj_{m+1}(\kappa_{in}r).$$

Outside the confining radius (r>R), the wavefunction is

$$a_m(r) = Pj_m(\kappa_{out}r) + Qy_m(\kappa_{out}r),$$
$$b_m(r) = sgn(E - V_{out})[Pj_{m+1}(\kappa_{out}r) + Qy_{m+1}(\kappa_{out}r)].$$

Here $j_m, y_m$ are the m-order Bessel functions of the first kind and the second kind respectively. F, P, Q are the coefficients of these Bessel functions. $\kappa_{in(out)} = |E - V_{in(out)}|/\hbar v_F$. The wavefunction is continuous at the confining boundary (r=R), thus we have the boundary condition

$$Fj_m(\kappa_{in}R) = Pj_m(\kappa_{out}R) + Qy_m(\kappa_{out}R),$$
$$sgn(E - V_{in})Fj_{m+1}(\kappa_{in}R) = sgn(E - V_{out})[Pj_{m+1}(\kappa_{out}R) + Qy_{m+1}(\kappa_{out}R)].$$

The other condition is that the wavefunction is normalized outside the confinement, $|P|^2 + |Q|^2 = 1$. From these, we can solve the values of F, P, Q and the wavefunction is obtained. Then the density of states is calculated by $\rho(E) = \frac{1}{2}|E - V_{out}||F|^2$. These equations are just the same as those in the theoretical work[2].

To simulate our condition, we set $R = 10$ nm, $V_{in} = 0.54$ eV, $V_{out} = 0$, $\hbar v_F = \frac{3}{2}|t|a = 0.6$ eV·nm. Here, the radius $R = 10$ nm is the same as that at our concise analytical figure (the right panel of Fig.1e) in main text, but the parameters $V_{in}$, $V_{out}$, $\hbar v_F$ are the same as these at our numerical figure (the left panel of Fig.1e) in main text because that they are not used in the concise analytical model. For the three orbital numbers m = 0, 1, 2, we get the density of states versus energy, as plotted in Fig. S1. The two peaks of the m = 0 curve correspond to the quasibound states with (m, n) = (0, 1) and (0, 2), for energy from high to low. The peaks of the m = 1 curve and m = 2 curve respectively correspond to states (1, 1) and (2, 1). Like the numerical and experimental results, the first two states are (0, 1) and (1, 1), and the next two states are (0, 2) and (2, 1) with close energy.

At each peak energy and according to the m value, we calculate the coefficients of Bessel functions $F$, $P$, $Q$, thus obtaining the wavefunction $\psi_{(0,1)}, \psi_{(1,1)}, \psi_{(0,2)}, \psi_{(2,1)}$. From this, we get consistent results with our concise analytical results (just setting A-sublattice Bessel function being zero at boundary). As for the states (0, 1) and (1, 1), we copy Extended data Fig. 1 for the circular confinement to the left panel and middle panel of Fig. S2, showing the numerical solutions and concise analytical solutions. For comparison, we also add the realistic analytical solutions of states (0, 1) and (1, 1) to the right panel of Fig. S2. The results obtained by these three methods are similar, which further demonstrates the robustness of our theoretical study.

Also, for the unhybridized states (0, 2) and (2, 1), we copy Fig. 1e of main text to the left and middle panels of Fig. S3, showing the numerical solutions and concise analytical solutions. For comparison, we also add the realistic analytical solutions of (0, 2) and (2, 1) to the right panel of Fig. S3. Note that these two states are close in energy and their energy broadening leads to their overlap. Thus, in the concise analytical solutions, we have used the combination of squared magnitudes of $\psi_{(0,2)}, \psi_{(2,1)}$, to better reflect the actual situation. Here for the realistic analytical solutions, we also plot the combination $0.6|\psi_{(0,2)}|^2 + 0.4|\psi_{(2,1)}|^2$ and $0.3|\psi_{(0,2)}|^2 + 0.7|\psi_{(2,1)}|^2$. The results from three methods are also very consistent.

Next, we study the hybridization induced by the elliptical perturbation. Just the same as the concise analytical derivation, the anisotropy induces the recombination between states (0, 2) and (2, 1). This forms the hybridized states $\psi_{sd+} = \psi_{(0,2)} + \alpha \psi_{(2,1)}$, $\psi_{sd-} = -\alpha \psi_{(0,2)} + \psi_{(2,1)}$. We copy the Fig. 1f of main text to the left and middle panels of Fig. S5, which shows the numerical and concise analytical results of hybridized states sd+ and sd-. In the concise analytical results, we have used the coefficient $\alpha = 0.7$. We apply the same coefficient to our realistic analytical solutions and plot the sd+ and sd- states, which are also shown in the right panel of Fig. S5. Again, the results obtained from three methods are very similar, consistently showing the hybridization features (one with $\theta$ shape and the other with rotated $\theta$ shape).

In summary, we can use the more realistic analytical method to reobtain the results of the confined states and their hybridization. These results are similar to both the numerical results and concise analytical results, indicating the robustness and the broad appeal of our theoretical study.

**IV. Discussion on the recombining coefficient.**

In general, the hybridization between states (0, 2) and (2, 1) can be written as:

$$\psi_{sd+} = \frac{1}{A}[\alpha_1 \psi_{(0,2)} + \alpha_2 \psi_{(2,1)}], \quad \psi_{sd-} = \frac{1}{A}[-\alpha_2^* \psi_{(0,2)} + \alpha_1^* \psi_{(2,1)}], \quad (S7)$$

with the normalization factor $A = \sqrt{|\alpha_1|^2 + |\alpha_2|^2}$ and coefficients $\alpha_1, \alpha_2$. In principle, the coefficients $\alpha_1, \alpha_2$ can be complex numbers. Eq. (S7) can be rewritten as:

$$\psi_{sd+} = \frac{\alpha_1}{A}[\psi_{(0,2)} + \alpha \psi_{(2,1)}], \quad \psi_{sd-} = \frac{\alpha_1^*}{A}[-\alpha^* \psi_{(0,2)} + \psi_{(2,1)}]. \quad (S8)$$

Here, $\alpha = \alpha_2/\alpha_1$ is also a complex number in principle. Note that the overall coefficients $\frac{\alpha_1}{A}, \frac{\alpha_1^*}{A}$ just modulate the overall values of wavefunctions. Thus, in the main text, for the sake of conciseness, we use the single-parameter formulas, i.e. the formulas within the square bracket in Eq. (S8), $\psi_{sd+} = \psi_{(0,2)} + \alpha\psi_{(2,1)}$, $\psi_{sd-} = -\alpha^*\psi_{(0,2)} + \psi_{(2,1)}$.

In fact, for the complex coefficient $\alpha$, its phase has the influence on patterns of hybridized states. As shown in Fig. S4, we plot the hybridized states with $\alpha = 0.7$ [Fig. S4a, the same as analytical results in Fig. 1f of main text] and $\alpha = 0.7i = 0.7e^{i\pi/2}$ [Fig. S4b]. One can find that the phase can rotate the hybridized states, changing the direction of their shapes.

In the main text, the minor axis and major axis of elliptical quantum dot (QD) are respectively along the x axis and y axis. This corresponds to a real $\alpha$. In this case the formulas can be rewritten as $\psi_{sd+} = \cos\theta\,\psi_{(0,2)} + \sin\theta\,\psi_{(2,1)}$, $\psi_{sd-} = -\sin\theta\,\psi_{(0,2)} + \cos\theta\,\psi_{(2,1)}$ with $\theta = \arctan\alpha$.

The use of Eq. (1) in main text is for both a more general physical picture and a concise single-parameter form. Although we have omitted the normalization factor, it does not affect the relative LDOS and thus also the figure plots.

**V. Numerical calculation for quasibound states.**

In numerical calculations, we obtain the LDOS from tight-binding model. The general Hamiltonian of graphene under the electrostatic potential is:[1]

$$H = \sum_{<i,j>} t a_i^\dagger a_j + \sum_i [\epsilon_0 + V(\boldsymbol{r}_i)] a_i^\dagger a_i.$$

Here $a_i$ ($a_i^\dagger$) denotes the annihilation (creation) operator at the discrete site $i$, $t = -2.8$ eV is the hopping energy between nearest-neighbor sites $<i,j>$, and $\epsilon_0 = 0$ is the onsite energy. We set that the QD leads to a potential $V(\boldsymbol{r}_i)$ as a function of the site coordinate $\boldsymbol{r}_i = (x_i, y_i)$ on the graphene.

The QD is set at the coordinate center (0, 0), and is described by a Coulomb potential with a cut-off radius. The potential can be circular or elliptical, and is uniformly written as[6]

$$V(\boldsymbol{r}_i) = \begin{cases} \beta, & \text{if } \tilde{r}_i < 1 \\ \dfrac{\beta}{\tilde{r}_i}, & \text{if } \tilde{r}_i > 1 \end{cases}$$

where $\tilde{r}_i = \sqrt{\dfrac{x_i^2}{r_1^2} + \dfrac{y_i^2}{r_2^2}}$ is the reduced distance between site $\boldsymbol{r}_i$ and the confining center. For $r_1 = r_2 = r_0$, the potential is circular. As we set $r_1 = r_0 - dr, r_2 = r_0 + dr$, the potential is deformed to be elliptical by the deformation length $dr$. The cut-off radius of the circular Coulomb potential is set as $r_0 = 6$ nm, and $dr$ is a variable. The height of the Coulomb potential is $\beta = 0.54$ eV. For the circular confinement, the Coulomb potential can be rewritten as $V(\boldsymbol{r}_i) = \hbar v_F \beta_0 / |\boldsymbol{r}_i|$ for $|\boldsymbol{r}_i| > r_0$, with the dimensionless Coulomb strength $\beta_0$ by $\beta_0 = \beta r_0 / \hbar v_F$.[6,8] Substituting $\hbar v_F = \dfrac{3}{2}|t|a$ with distance between nearest carbon atoms $a = 0.142$ nm, we obtain $\beta_0 = 5.4$. This value is much larger than the critical value $\beta_{0,c} = 0.5$ for the emergence of atomic collapse state (ACS)[6,8,9], thus the system is in the supercritical regime and ACSs can appear under such a Coulomb potential.

To eliminate the finite-size effect, the calculations are performed in a large 80 nm×80 nm square graphene centered at coordinate origin (0, 0), which is also the center point of the QD potential. Based on the kernel polynomial method[10], we use an open source code package *pybinding*[11] to calculate the local density of states $\text{LDOS}(\boldsymbol{r_i})$ of any discretized graphene site $\boldsymbol{r_i}$. The energy broadening is set to be 0.01 eV. In order to obtain the LDOS under continuously changed coordinates $\boldsymbol{r}$, we make a weighted sum of discretized $\text{LDOS}(\boldsymbol{r_i})$

$$\text{LDOS}(\boldsymbol{r}) = \sum_i e^{-\frac{(r-r_i)^2}{\gamma^2}} \text{LDOS}(\boldsymbol{r_i}).$$

This represents that $\text{LDOS}(\boldsymbol{r})$ is mainly contributed by the sites nearby $\boldsymbol{r} = (x,y)$ within the distance $\gamma = 0.25$ nm.

For $dr$ = 0, 0.15 nm, …, 1.20 nm corresponding to scaled deformation $dr' = 0, 0.025, …, 0.200$, we calculate the space-energy LDOS along the horizonal minor axis with $y = 0$. The LDOS for 5 representative $dr' = 0, 0.05, 0.10, 0.15, 0.20$ are shown in Extended data Figs. 10a1-10a5. As shown in Extended data Fig. 10a1, for the circular QD with $dr = 0$, the quasibound states indicate (m, n)= (0, 1), (1, 1), (0, 2), and (2, 1), with energy from high to low. Respectively, these states are recorded as s1, p1, s2 and d1, which are indicated by red, yellow, green and blue bars. s1 and s2 are ACSs[6,12], while p1 and d1 are whispering gallery modes (WGMs)[3,6]. When $dr$ increases, the circular QD is deformed to be elliptical, the states s2 and d1 hybridize and the energy splits, which is shown by the color bars in Extended data Figs. 10a2-10a5. To extract the energy of quasibound states, we use the two LDOS-energy curves at coordinates $(x,y) =$(0,0) and (5 nm, 0). Because the states (0, 1) and (0, 2) concentrate in the center, the energy of s1, s2 and sd+ are extracted from the peaks of curves at (0, 0). The states (1, 1) and (2, 1) concentrate in the edge, and the energy of p1, d1 and sd- are extracted from the peaks of curves at (5 nm, 0). Based on the quasibound state energies, the scaled energy $E', E'_3, E'_4$ are obtained. By doing so, the scaled energy splitting versus $dr'$ is shown in Fig. 3e. For each $dr'$ and according to the extracted energy of quasibound states, the spatial LDOS distribution of quasibound states can also be calculated. By doing this, the numerical results in Extended data Figs. 10b1-10b5, 10c1-10c5 are correspondingly obtained. The other numerical results are calculated by only changing $dr$ (equivalent to changing $dr'$), with other parameters unchanged: The numerical LDOS distributions in Figs. 1e, 1f are obtained by setting $dr' = 0, 0.16$ respectively. The numerical results in Fig. 3 are obtained according to the corresponding $dr'$.

Because in experiments it is very hard to fabricate a perfectly circular QD, the QDs with $dr' < 0.05$ in experiments are indeed a bit deviated from perfect circle. However, the tiny deformation has almost no influence on the results, and there is nearly no hybridization between states s2 (0, 2) and d1 (2, 1). For illustration, in Fig. S11, we show the numerically calculated LDOS of s1, p1, s2, d1 for a small $dr' = 0.025$, which are almost the same as the results for $dr' = 0$ (Extended data Fig. 1 for s1, p1, Fig. 1e for s2, d1). Therefore, the almost circular QDs almost does not induce hybridization and give a concise physical picture for unhybridized cases.

**VI. Discussion on STM tip potential.**

To minimize the impact of the STM tip[13] on the QD structures during measurements, we use a tungsten tip with a work function closely matching that of graphene[14,15]. This alignment leads to a relatively small tip-induced potential[16]. Additionally, the spatial extent of this potential is large, exceeding 40 nm or even more[3,17]. In contrast, the QDs under investigation typically have a radius of around 10 nm, leading to a narrow and high potential barrier, with a potential difference of approximately 500 meV between the regions on and outside the QD[6]. Due to the broad spatial range of the tip-induced potential relative to the compact size of the QD, this potential behaves as a slowly varying and relatively small field over the area of the QD. Consequently, it is unlikely to induce any localized effects on the electronic properties within the graphene QD structure, thereby facilitating accurate measurements of the intrinsic electronic states.

We also numerically calculate the influence of STM tips. We set the STM tip additionally induces a Lorentz potential around its center $(x_{STM}, 0)$.

$$V_{STM} = \beta_{STM} \frac{r_{STM}^2}{r_{STM}^2 + (x - x_{STM})^2 + y^2} - \beta_{STM}.$$

Here $\beta_{STM}$ is the potential height, and $r_{STM} = 40$ nm is the influence range of the STM tip. In the potential, we subtract $\beta_{STM}$ to reset the potential at STM center being zero. We set the STM tip lies at $x_{STM} = 5$ nm, to study the QD symmetry is broken to what extent. The results are shown in Fig. S6.

From Fig. 3a of main text, we directly copy the space-energy LDOS maps for $dr' = 0.04, 0.18$ to Figs. S6a1 and S6b1, respectively. As shown in Figs. S6a2-S6a4 and S6b2-S6b4, we add the STM potential $\beta_{STM}$ and find that the LDOS maps are almost unaffected by the STM potential, even for a large value $\beta_{STM} = 0.4$ eV.

Therefore, when measuring, our tungsten STM tip has almost no influence on the confined states. The measured features are just the properties of confined states themselves.

## VII. Experimental energy extraction of confined states.

To extract the energy values $E_2$, $E_3$, $E_4$ and their corresponding $E_i'$ ($i = 3, 4$), we follow a systematic method based on the localization of states. The sd+ state is primarily localized at the center of the QD so we extract the energy value $E_3$ from the spectra at the center. In contrast, the (1, 1) and sd- states are mainly localized at the edges of the QD, allowing us to obtain $E_2$ and $E_4$ from the spectra at the edges. The energy scale is then defined as $E_i' = \frac{E_i - (E_3 + E_4)/2}{E_2 - (E_3 + E_4)/2}$ ($i = 3,4$), leading to $E_3'$ and $E_4'$. The detailed procedure is as follows:

1. **Peak Extraction**: First, we take the negative second derivative of the d$I$/d$V$ spectra to obtain clearer peak information (see Figs. S7a to S7c). The data was smoothed during the calculation of the negative second derivative to enhance feature clarity and reduce noise.

2. **Data Collection**: We collect d$I$/d$V$ spectra within a 1 nm range around the center of the QD (Fig. S7d) and extract all peaks corresponding to $E_3$ (Fig. S7e), calculating the average value $\bar{E}_3$ and standard deviation $\sigma_3$. Similarly, we extract peaks for $E_2$ and $E_4$ from the spectra collected within a 1 nm range on both sides of the QD, approximately 2 nm from its boundary (Fig. S7d), also calculating the average values and standard deviations.

3. **Energy Scale Calculation**: Using the formula $E_i' = \frac{E_i - (E_3 + E_4)/2}{E_2 - (E_3 + E_4)/2}$ ($i = 3,4$), we can calculate the energy-scaled values $E_3'$ and $E_4'$. During this process, we substitute the corresponding $i$ values ($i = 3,4$) and then apply the error propagation formula:

$$\sigma_i' = \sqrt{\left(\frac{\partial E_i'}{\partial E_2}\sigma_2\right)^2 + \left(\frac{\partial E_i'}{\partial E_3}\sigma_3\right)^2 + \left(\frac{\partial E_i'}{\partial E_4}\sigma_4\right)^2}$$

to obtain the corresponding error values for $E'_3$ and $E'_4$.

## Supplementary References

# Supplementary Figures

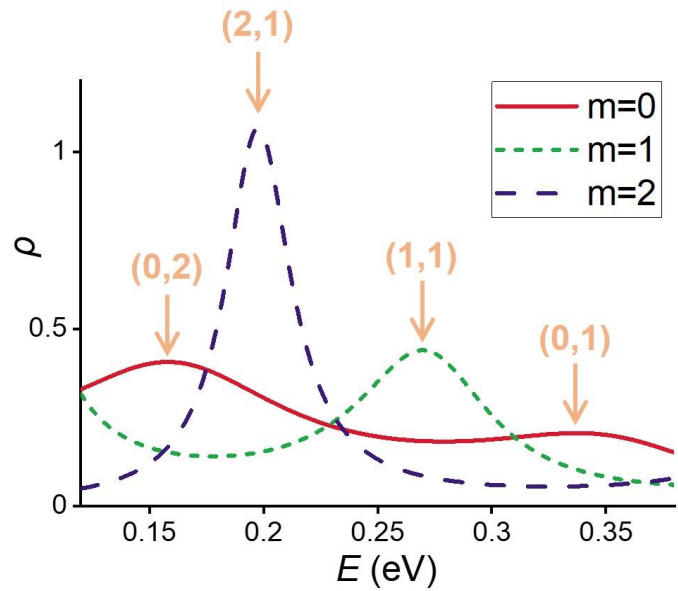

**FIG. S1. The density of states from the realistic analytical method.** Different curves correspond to the density of states of different orbital numbers m. The arrows mark the energy positions of the first four confined states.

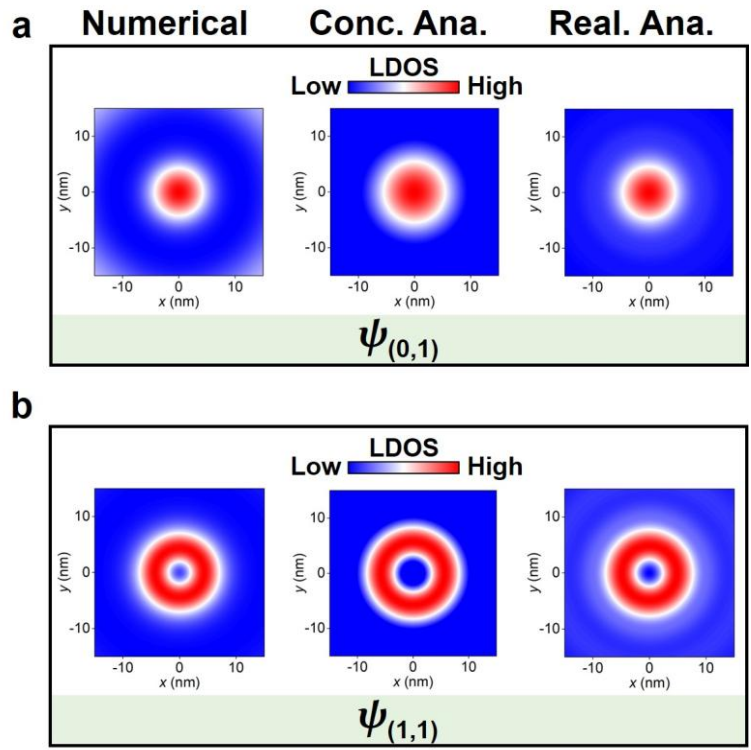

**FIG. S2. The confined states (0, 1) and (1, 1) obtained by three methods. a.** The spatial distribution of state (0, 1). **b.** The spatial distribution of state (1, 1). The left, middle, and right panels are the results from the numerical calculations, concise analytical derivations, and realistic analytical derivations, respectively.

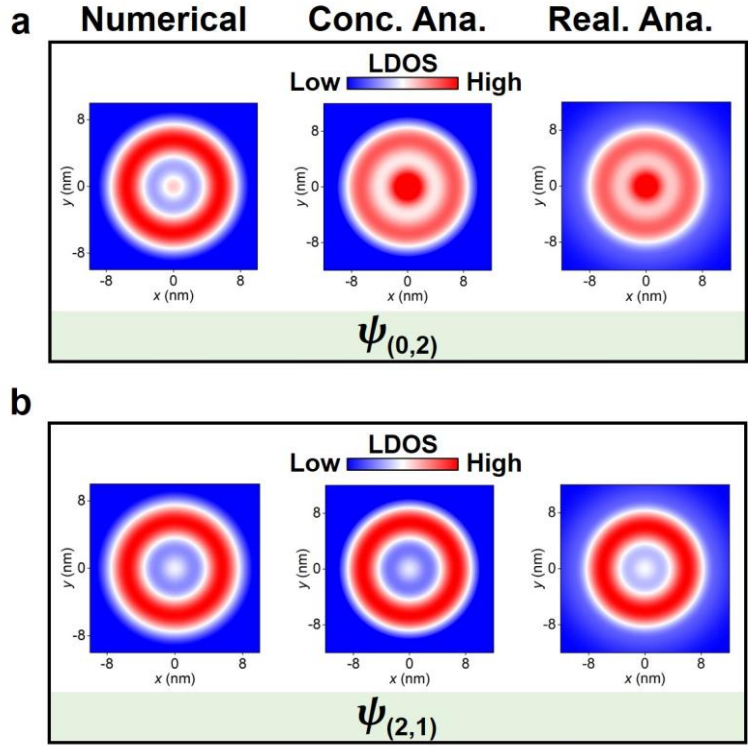

**FIG. S3. The unhybridized states (0, 2) and (2, 1) obtained by three methods. a.** The spatial distribution of state (0, 2). **b.** The spatial distribution of state (2, 1). The left, middle, and right panels are the results from the numerical calculations, concise analytical derivations, and realistic analytical derivations, respectively.

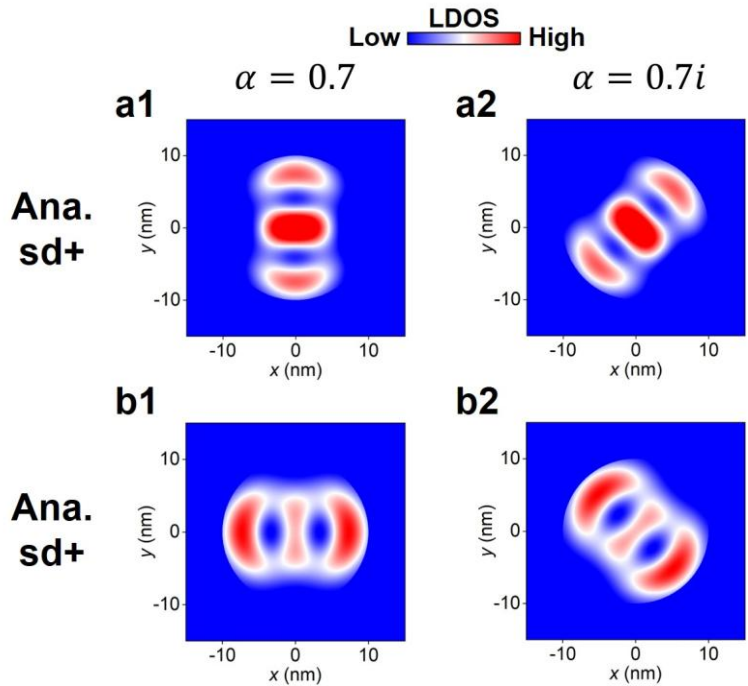

**FIG. S4. The influence of phase of $\alpha$ on hybridized states. a.** The hybridized states for a real coefficient $\alpha = 0.7$. **b.** The hybridized states for an imaginary coefficient $\alpha = 0.7i$.

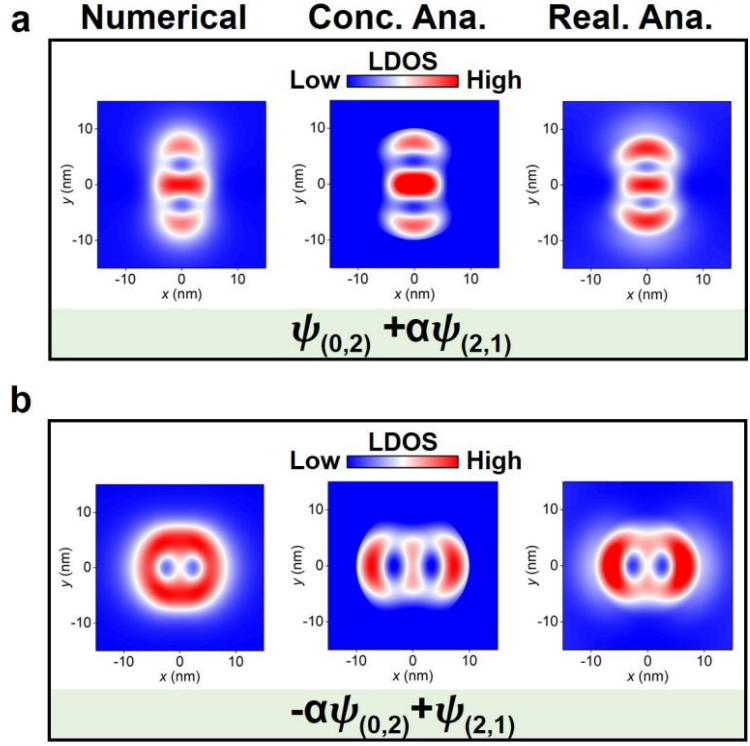

**FIG. S5. The hybridized states sd+ and sd- obtained by three methods. a.** The spatial distribution of state sd+. **b.** The spatial distribution of state sd-. The left, middle, and right panels are the results from the numerical calculations, concise analytical derivations, and realistic analytical derivations, respectively.

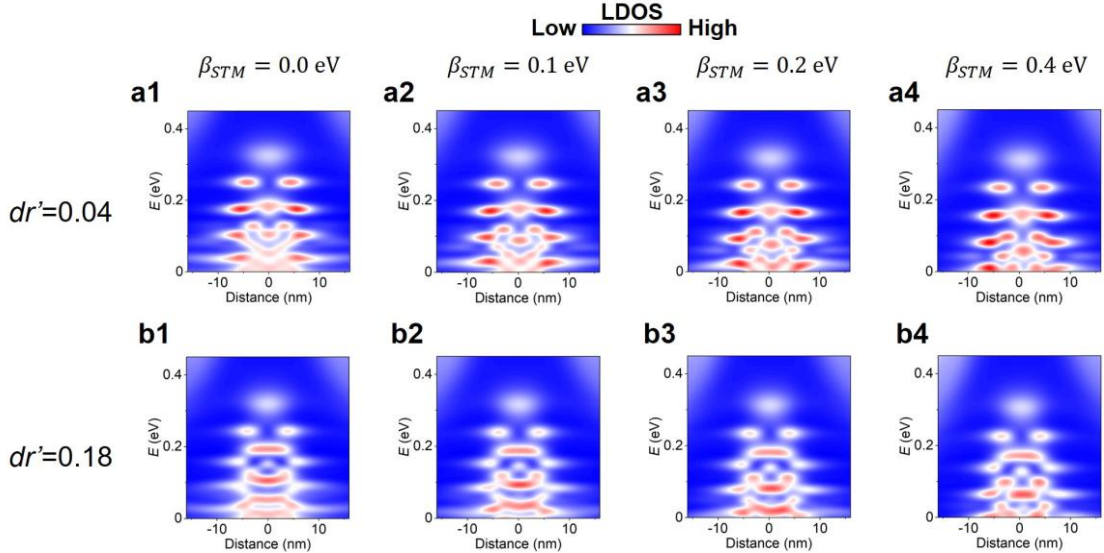

**FIG. S6. The weak influence of STM potentials shown by numerical calculations.** From leftmost panel to rightmost panel, we show the space-energy LDOS for different STM potential values $\beta_{STM} = 0.0, 0.1, 0.2, 0.4$ eV. **a1-a4.** The influence on the result with $dr' = 0.04$. **b1-b4.** The influence on the result with $dr' = 0.18$. The leftmost panels are the same as those in Fig. 3a with $dr' = 0.04$ and $dr' = 0.18$ in the main text. The other parameters are the same as those in Fig. 3a of main text.

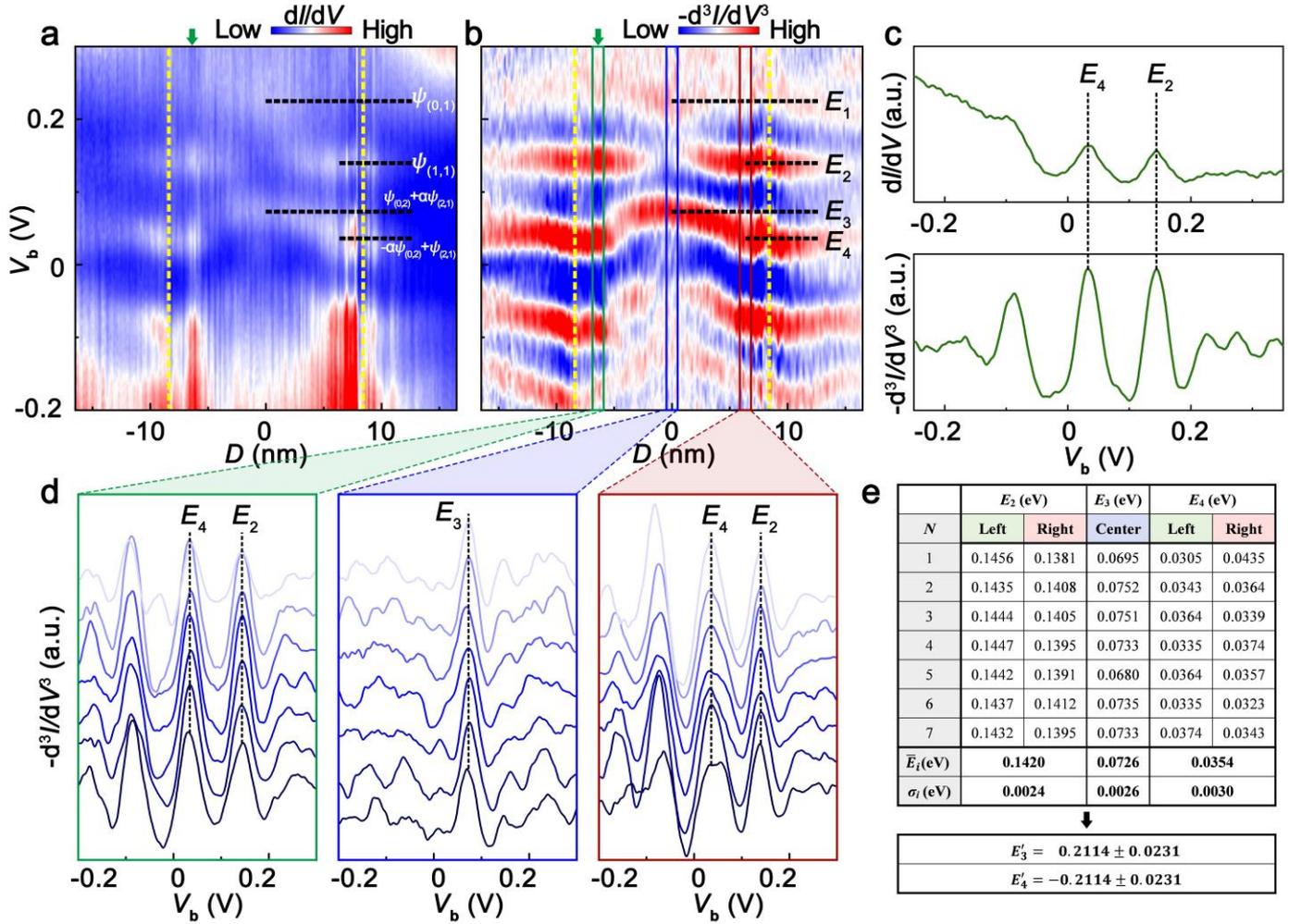

**FIG. S7. Energy extraction and scaling process for QD states. a.** The d$I$/d$V$ spectroscopic map along the minor axis of the elliptical QD with $dr' = 0.18$. Orbital states can be observed in the QD. The first four states are labelled by wavefunctions and black dashed lines. The two yellow dashed lines mark the size of the QD. **b.** Experimental $-d^3I/dV^3$ spectroscopic map of the elliptical QD, revealing clearer orbital states. For simplicity, the first four states are represented as $E_i$ (where $i = 1,2,3,4$). **c.** The upper and lower panels correspond to the line cuts at the positions marked by green arrows in panels a and b, respectively. By taking the negative second derivative, the peak positions become clearer. **d.** $-d^3I/dV^3$ spectra corresponding to the three colored boxed regions in panel b. The centers of the green and red regions are positioned 2 nm from the boundary of the QD, while the blue region is located at the center of the QD. Each region spans 1 nm in length. The data are offset along the y-axis for clarity. **e.** Top panel: The table displays the extracted energy values of the confined states from panel d, along with their respective averages and standard deviations. Bottom panel: The scaled values of $E'_3$ and $E'_4$, obtained using the formula $E'_i = \frac{E_i - (E_3 + E_4)/2}{E_2 - (E_3 + E_4)/2}$ ($i = 3,4$), along with the corresponding uncertainties calculated using the error propagation formula.

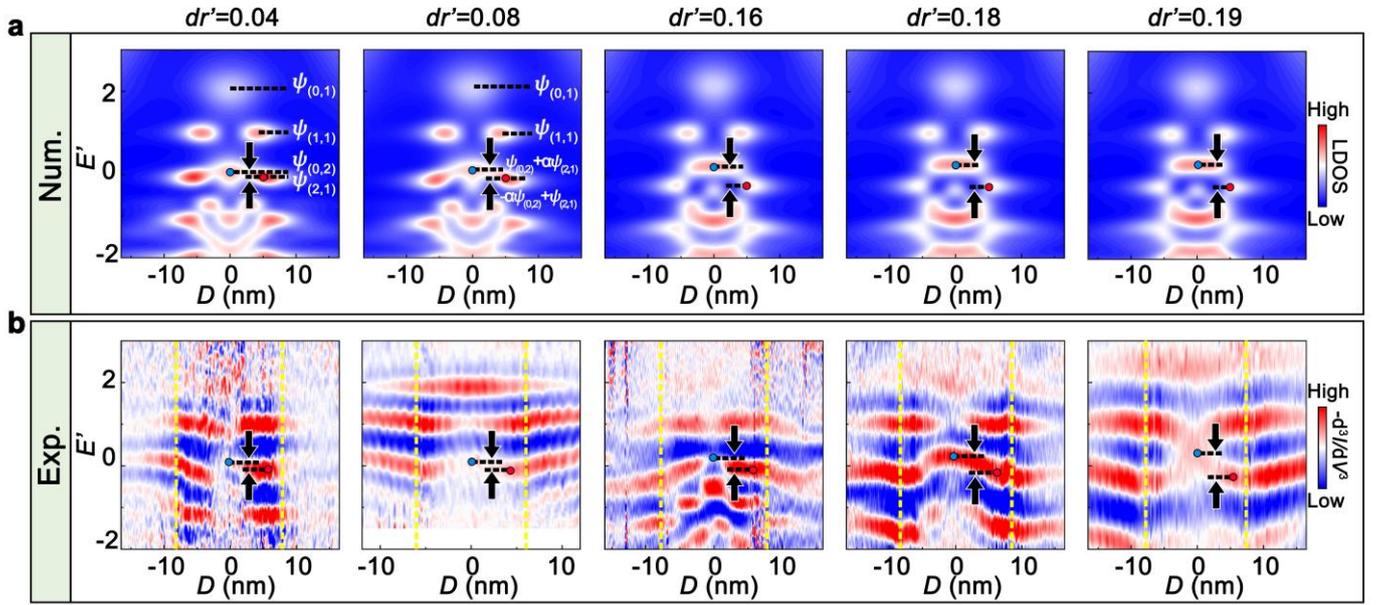

**FIG. S8. More detailed evolution of space-energy maps. a.** Similar to Fig. 3a, but with an additional map where dr' = 0.08. **b.** Similar to Fig. 3b, but with an additional map where dr' = 0.08.

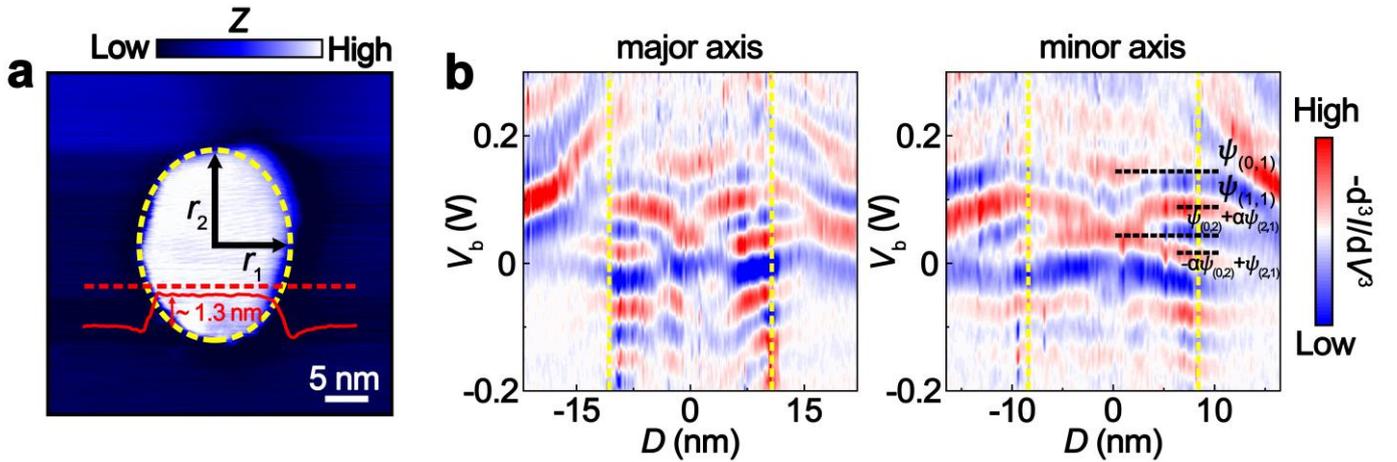

**FIG. S9. Hybridized orbital states under an elliptical confinement with $dr' = 0.12$. a.** A STM image ($V_b$ = 500 mV, $I$ = 100 pA) of the elliptical QD with anisotropy degree $dr' = 0.12$ embedded in the graphene/WSe$_2$ heterostructure. The minor radius $r_1$ is approximately 8.4 nm, and the major radius $r_2$ is approximately 10.7 nm. The height profile along the red dashed line is shown with solid red line. **b.** The $-d^3I/dV^3$ spectroscopic maps versus the spatial position along the major axis (left panel) and minor axis (right panel) of the elliptical QD, respectively. Orbital states can be clearly observed in the QD. The first four states are labelled by wavefunctions and black dashed lines in the right panel. The yellow dashed lines mark the size of the QD.

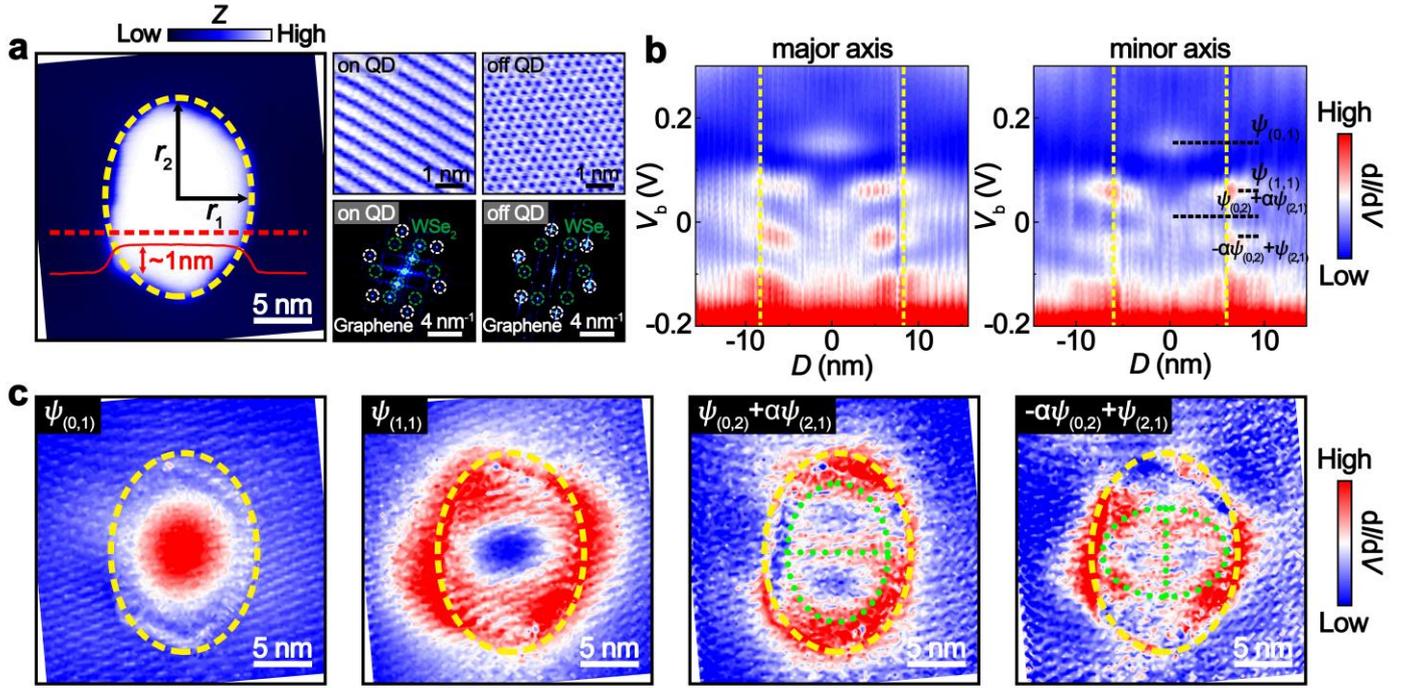

**FIG. S10. Hybridized orbital states under an elliptical confinement with $dr' = 0.16$. a.** Left panel: A STM image ($V_b$ = 400 mV, $I$ = 150 pA) of the elliptical QD with anisotropy degree $dr' = 0.16$ embedded in the graphene/WSe$_2$ heterostructure. The minor radius $r_1$ is approximately 6.0 nm, and the major radius $r_2$ is approximately 8.3 nm. The height profile along the red dashed line is shown with solid red line. Top right panels: Atomic-resolved STM image on (1T' phase) and off (2H phase) the QD, respectively. Bottom right panels: The FFT image obtained from the STM image on and off the QD, respectively. The white and green circles show reciprocal lattices of graphene and WSe$_2$, respectively. The unlabeled bright spots correspond to the reciprocal moiré superlattices and higher-order scattering. **b.** The d$I$/d$V$ spectroscopic maps versus the spatial position along the major axis (left panel) and minor axis (right panel) of the elliptical QD, respectively. Orbital states can be clearly observed in the QD. The first four states are labelled by wavefunctions and black dashed lines in the right panel. The yellow dashed lines mark the size of the QD. **c.** d$I$/d$V$ maps of different orbital states. For an elliptical confinement, the anisotropy of confining potential results in orbital hybridization between the s-orbital and d-orbital states, giving rise to new states sd+ $\psi_{(0,2)} + \alpha\psi_{(2,1)}$ and sd- $-\alpha\psi_{(0,2)} + \psi_{(2,1)}$, which exhibit $\theta$-shaped and rotated $\theta$-shaped features marked by the green dotted lines, respectively. The yellow dashed lines show outlines of the elliptical QD.

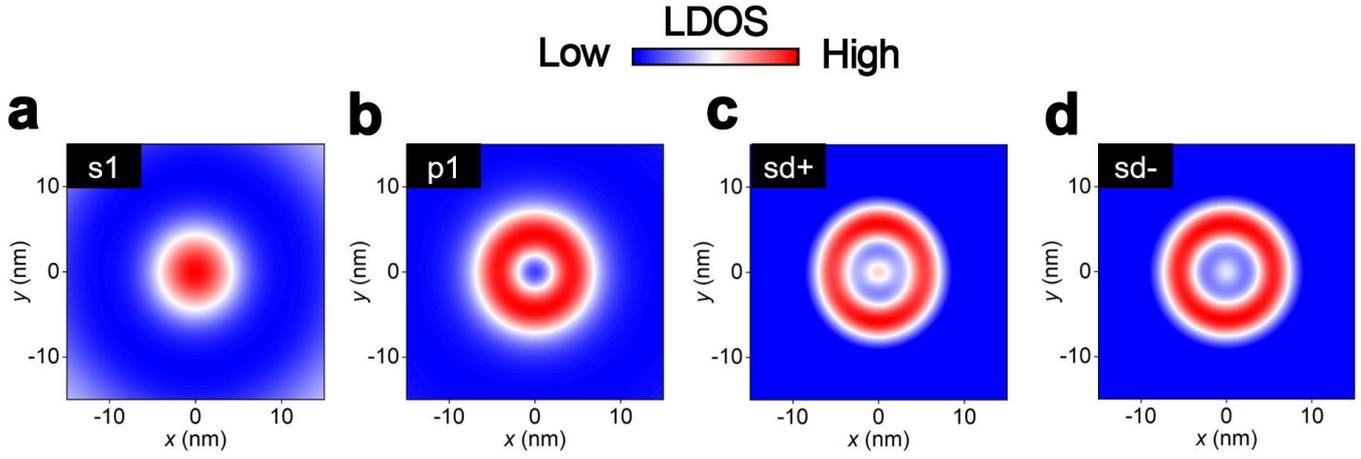

**FIG. S11. Numerically calculated quasibound states for a very small $dr' = 0.025$. a to d.** The LDOS distributions of quasibound states s1 (0,1), p1 (1,1), s2 (0,2), and d1 (2,1) for a QD slightly deformed from a circle shape. The parameters except $dr' = 0.025$ are the same as those in numerical calculations for the circle QD in Extended data Fig. 1. The small deformation $dr'$ does not visibly affect the results.

## Supplementary Table

**Table S1.** The absolute energy values corresponding to Fig. 3e in main text.

| $dr'$ | $E_2$ (eV) | $E_3$ (eV) | $E_4$ (eV) |
|---|---|---|---|
| 0.034±0.003 | 0.1838±0.0099 | 0.0915±0.0017 | 0.0718±0.0069 |
| 0.037±0.007 | 0.0945±0.0042 | -0.0022±0.0075 | -0.0171±0.0061 |
| 0.073±0.002 | 0.0821±0.0020 | 0.0417±0.0014 | 0.0308±0.0020 |
| 0.081±0.006 | 0.0377±0.0021 | -0.0678±0.0076 | -0.0891±0.0033 |
| 0.124±0.006 | 0.0906±0.0030 | 0.0442±0.0024 | 0.0256±0.0103 |
| 0.157±0.004 | 0.0744±0.0045 | -0.0022±0.0018 | -0.0288±0.0077 |
| 0.163±0.004 | 0.0590±0.0024 | 0.0109±0.0033 | -0.0293±0.0033 |
| 0.167±0.008 | 0.1078±0.0017 | 0.0578±0.0034 | 0.0291±0.0034 |
| 0.181±0.008 | 0.1420±0.0024 | 0.0726±0.0026 | 0.0354±0.0030 |
| 0.193±0.008 | 0.1105±0.0032 | 0.0716±0.0035 | 0.0368±0.0035 |
| 0.214±0.007 | 0.1193±0.0023 | 0.0730±0.0020 | 0.0446±0.0075 |

**Supplementary Codes**

The codes for the analytical studies are run via MATLAB.

The analytical results of states (0, 1) and (1, 1) are obtained by running the code below:

```
clear all;
l0 = 0; l1 = 1; l2 = 2; l3 = 3; alpha = 0.0;
X = -15.02:0.04:15.02; Y = -15.02:0.04:15.02;
r0 = 10; bessel01 = 2.405; bessel02 = 5.520; bessel21 = 5.136; bessel11 = 3.832;
k01 = bessel01/r0; k02 = bessel02/r0; k1 = bessel11/r0; k2 = bessel21/r0;
THETA = zeros(length(Y),length(X));
PSI1 = zeros(length(Y),length(X));
PSI2 = zeros(length(Y),length(X));
for indx = 1:length(X)
    x = X(indx);
    R = (x^2+Y.^2).^0.5; Theta = atan(Y./x) - sign(x)*pi/2 + pi/2;
    THETA(:,indx) = Theta;
    b0 = besselj(l0,k01*R).* exp(1j*l0*Theta);
    b1 = -1j*besselj(l1,k01*R).* exp(1j*l1*Theta);
    b2 = besselj(l1,k1*R).* exp(1j*l1*Theta);
    b3 = -1j*besselj(l2,k1*R).* exp(1j*l2*Theta);
    PSI1(:,indx) = ((abs(   (1)*b0+(alpha)*b2   )).^2+(abs(   (1)*b1+(alpha)*b3   )).^2).*(sign(abs(r0)^2-R.^2)+1)/2;
    PSI2(:,indx)   =   ((abs(     (-alpha)*b0+(1)*b2     )).^2+(abs(     (-alpha)*b1+(1)*b3   )).^2).*(sign(abs(r0)^2-R.^2)+1)/2;
end
subplot(1,2,1);pcolor(X,Y,PSI1);colorbar;shading interp;
subplot(1,2,2);pcolor(X,Y,PSI2);colorbar;shading interp;
```

The analytical results of states (0, 2) and (2, 1), along with their hybridization (regulated by alpha in red), are obtained by running the code below:

```
clear all;
l0 = 0; l1 = 1; l2 = 2; l3 = 3; alpha = 0.7*exp(1j*0.0*pi);
X = -15.02:0.04:15.02; Y = -15.02:0.04:15.02;
r0 = 10; bessel01 = 2.405; bessel02 = 5.520; bessel21 = 5.136; bessel11 = 3.832;
k01 = bessel01/r0; k02 = bessel02/r0; k1 = bessel11/r0; k2 = bessel21/r0;
THETA = zeros(length(Y),length(X));
PSI1 = zeros(length(Y),length(X));
PSI2 = zeros(length(Y),length(X));
for indx = 1:length(X)
    x = X(indx);
    R = (x^2+Y.^2).^0.5; Theta = atan(Y./x) - sign(x)*pi/2 + pi/2;
    THETA(:,indx) = Theta;
    b0 = besselj(l0,k02*R).* exp(1j*l0*Theta);
    b1 = -1j*besselj(l1,k02*R).* exp(1j*l1*Theta);
    b2 = besselj(l2,k2*R).* exp(1j*l2*Theta);
    b3 = -1j*besselj(l3,k2*R).* exp(1j*l3*Theta);
    PSI1(:,indx) = ((abs(  (1)*b0+(alpha)*b2  )).^2+(abs(  (1)*b1+(alpha)*b3  )).^2).*(sign(abs(r0)^2-R.^2)+1)/2;
    PSI2(:,indx) = ((abs(  (-alpha')*b0+(1)*b2  )).^2+(abs(  (-alpha')*b1+(1)*b3  )).^2).*(sign(abs(r0)^2-R.^2)+1)/2;
end
subplot(1,2,1);pcolor(X,Y,PSI1);colorbar;shading interp;
subplot(1,2,2);pcolor(X,Y,PSI2);colorbar;shading interp;
```

The codes for the numerical studies are run via python and the package *pybinding*.

For the numerical space-energy maps of LDOS, they are obtained by running the code below. The dr' corresponds to dr0 in red and can be regulated.

```python
import pybinding as pb
import numpy as np
import matplotlib.pyplot as plt
from scipy import io
pb.pltutils.use_style()
from pybinding.repository import graphene
from scipy.spatial import cKDTree
import math

c0 = 0.142 * 3

t0 = -2.8
e0 = -0.0

theta_p = 0*math.pi

x_p = 0
y_p = 0
r_0 = 6
dr0 = 0.100
r_1 = r_0 * (1 - dr0)
r_2 = r_0 * (1 + dr0)
beta = 0.54

zhankuan = 0.25
cutoff = 3*zhankuan

L_x_region = 44
L_y_region = 3
theta_region = 0.00*math.pi
site1 = [-L_x_region/2, -L_y_region/2]
site2 = [-L_x_region/2,  L_y_region/2]
site3 = [ L_x_region/2,  L_y_region/2]
site4 = [ L_x_region/2, -L_y_region/2]
xuanzhuan = [[math.cos(theta_region),-math.sin(theta_region)],[math.sin(theta_region),math.cos(theta_region)]]
site1xz = np.dot(xuanzhuan, site1)
```

```python
site2xz = np.dot(xuanzhuan, site2)
site3xz = np.dot(xuanzhuan, site3)
site4xz = np.dot(xuanzhuan, site4)
region = pb.Polygon([site1xz, site2xz, site3xz, site4xz])

def potential():
    @pb.onsite_energy_modifier
    def f(energy,x,y):
        R=np.sqrt((x*math.cos(theta_p)+y*math.sin(theta_p)-x_p)**2/r_1**2+(-x*math.sin(theta_p)+y*math.cos(theta_p)-y_p)**2/r_2**2)
        V1=np.zeros((len(R)))
        print(len(V1))
        V1[R <= 1] += beta
        V1[R > 1] += beta/R[R > 1]
        print(len(R), sum(R <= 1),sum(R > 1))
        return V1+energy+e0
    return f

shape = pb.rectangle(80,80)
model = pb.Model(
        graphene.monolayer(t=t0),
        shape,
        potential()
)

kpm=pb.kpm(model)
Enum = np.linspace(-0.2, 0.6, 1601)
spatial_ldos = kpm.calc_spatial_ldos(energy=Enum, broadening=0.01, shape=region)

ALLDOS = spatial_ldos.data.T
XYZ = np.array(spatial_ldos.structure.xyz[:,0:2])

R_DOS = np.array([np.linspace(-20, 20, 401)*math.cos(theta_region),np.linspace(-20, 20, 401)*math.sin(theta_region)]).T

tree = cKDTree(XYZ)
results = tree.query_ball_point(R_DOS, cutoff)

import scipy.sparse as ss
row = []
```

```python
col = []
data = []
for r_n in range(R_DOS.shape[0]):
    result = results[r_n]
    length = np.array(result).shape[0]
    col=np.append(col,result)
    row=np.append(row,r_n*np.mat(np.ones((1,length))))
    for ind_n in range(length):
        distance = np.linalg.norm(R_DOS[r_n,:]-XYZ[result[ind_n],:])
        xishu = math.exp( -distance**2/zhankuan**2 )
        data.append(xishu)
TRANS = ss.coo_matrix((data,(row,col)),shape=(R_DOS.shape[0],XYZ.shape[0]))
LDOS = TRANS @ ALLDOS

io.savemat('LDOS_1C_1D_theta0r60dr100.mat', {'LDOS':LDOS.T,'R_DOS':R_DOS,'Enum':Enum})
```

From the space-energy maps, one can extract the energy values for quasibound states. Then according to the energy values, one can obtain the spatial distributions of quasibound states via the code below. The energy values are in green and the dr0 is in red.

```python
import pybinding as pb
import numpy as np
import matplotlib.pyplot as plt
from scipy import io
pb.pltutils.use_style()
from pybinding.repository import graphene
from scipy.spatial import cKDTree
import math

c0 = 0.142 * 3

t0 = -2.8
e0 = -0.0

theta_p = 0*math.pi

x_p = 0
y_p = 0
r_0 = 6
dr0 = 0.100
r_1 = r_0 * (1 - dr0)
r_2 = r_0 * (1 + dr0)
beta = 0.54

zhankuan = 0.25
cutoff = 3*zhankuan

def potential():
    @pb.onsite_energy_modifier
    def f(energy,x,y):
        R=np.sqrt((x*math.cos(theta_p)+y*math.sin(theta_p)-x_p)**2/r_1**2+(-x*math.sin(theta_p)+y*math.cos(theta_p)-y_p)**2/r_2**2)
        V1=np.zeros((len(R)))
        print(len(V1))
        V1[R <= 1] += beta
        V1[R > 1] += beta/R[R > 1]
        print(len(R), sum(R <= 1),sum(R > 1))
```

```
            return V1+energy+e0
        return f

shape = pb.rectangle(80,80)
model = pb.Model(
            graphene.monolayer(t=t0),
            shape,
            potential()
)

kpm=pb.kpm(model)
Enum = np.array([0.3225, 0.2485, 0.1905, 0.1730 ])
spatial_ldos = kpm.calc_spatial_ldos(energy=Enum, broadening=0.01, shape=pb.rectangle(32,32))

ALLDOS = spatial_ldos.data.T
XYZ = np.array(spatial_ldos.structure.xyz[:,0:2])

x_plt, y_plt = np.mgrid[-15:15.25:0.25,-15:15.25:0.25]
R_DOS = np.c_[x_plt.ravel(), y_plt.ravel()]
tree = cKDTree(XYZ)
results = tree.query_ball_point(R_DOS, cutoff)

import scipy.sparse as ss
row = []
col = []
data = []
for r_n in range(R_DOS.shape[0]):
    result = results[r_n]
    length = np.array(result).shape[0]
    col=np.append(col,result)
    row=np.append(row,r_n*np.mat(np.ones((1,length))))
    for ind_n in range(length):
        distance = np.linalg.norm(R_DOS[r_n,:]-XYZ[result[ind_n],:])
        xishu = math.exp( -distance**2/zhankuan**2 )
        data.append(xishu)
TRANS = ss.coo_matrix((data,(row,col)),shape=(R_DOS.shape[0],XYZ.shape[0]))
LDOS = TRANS @ ALLDOS

io.savemat('LDOS_1C_2D_theta0r60dr100.mat', {'LDOS':LDOS.T,'R_DOS':R_DOS,'Enum':Enum})
```